\documentclass[12pt]{article}
\usepackage{graphicx} % Required for inserting images
\usepackage[textwidth=12em,textsize=small]{todonotes}
\usepackage[utf8]{inputenc}
\usepackage{mathbbol}
\usepackage{bm}
\usepackage{url}
\usepackage{amsmath}
\usepackage{natbib}
\usepackage[utf8]{inputenc}
\usepackage{xcolor}
\usepackage{graphicx}
\usepackage{enumerate}
\usepackage{amsthm}
\usepackage{amsmath}
\usepackage{amssymb}
\usepackage{marginnote}
\usepackage{mathtools}
\usepackage{float}

\usepackage{booktabs} %Nice tables
\usepackage{changepage}
\usepackage{color}
\usepackage{setspace}
\usepackage{lscape}
\usepackage{algorithm}
\usepackage{algpseudocode}

\usepackage[top=0.7in, bottom=0.7in, left=1in, right=1in]{geometry}

\begin{document}
\begin{center}
{\Large The Signal is Not Flushed Away: Inferring the Effective Reproduction Number From Wastewater Data in Small Populations 
}\\ \ \\

% List of authors
Isaac H. Goldstein$^1$, 
Daniel M. Parker$^2$, 
Sunny Jiang$^3$,
Aiswarya Rani Pappu$^3$,
Volodymyr M. Minin$^4$\\
%% Author addresses
$^1$Department of Statistics, Stanford University \\
$^2$Department of Population Health and Disease Prevention; Department of Epidemiology and Biostatistics, University of California, Irvine \\
$^3$Department of Civil and Environmental Engineering; Department of Ecology and Evolutionary Biology, University of California, Irvine\\
$^4$Department of Statistics, University of California, Irvine \\[2pt]
%% E-mail address for correspondence
\end{center}
\begin{abstract}{
The effective reproduction number is an important descriptor of an infectious disease epidemic. 
In small populations, ideally we would estimate the effective reproduction number using a Markov Jump Process (MJP) model of the spread of infectious disease, but in practice this is computationally challenging. 
We propose a computationally tractable approximation to an MJP which tracks only latent and infectious individuals, the EI model, an MJP where the time-varying immigration rate into the E compartment is equal to the product of the proportion of susceptibles in the population and the transmission rate. 
We use an analogue of the central limit theorem for MJPs to approximate transition densities as normal, which makes Bayesian computation tractable. 
Using simulated pathogen RNA concentrations collected from wastewater data, we demonstrate the advantages of our stochastic model over its deterministic counterpart for the purpose of estimating effective reproduction number dynamics, and compare against a state of the art method.  
We apply our new model to inference of changes in the effective reproduction number of SARS-CoV-2 in several college campus communities that were put under wastewater pathogen surveillance in 2022. 
}
\end{abstract}

\section{Introduction}
% talk about the problems of small populations, stemr, pomp, etc.
Pathogen genome concentrations collected from wastewater (henceforth referred to as wastewater data) provide information about the number of currently infected and recently recovered individuals in an infectious disease outbreak, and thus are a potential source of data when modeling an epidemic \citep{hillary2020wastewater,polo2020making}. 
There are many examples of studies that correlate wastewater data with other data sources, such as counts of new cases (for example \citet{song2021detection}), as well as a growing number of instances of wastewater data being incorporated into epidemiological statistical models \citep{lison2024improving,miyazawa2024wastewater,jin2024combining,nourbakhsh2022wastewater,goldstein2024semiparametric,morvan2022analysis}.
To our knowledge, existing methods have largely focused on large populations, using data harvested from wastewater treatment plants which often process wastewater from hundreds of thousands of individuals. 
However, epidemiological modeling of small populations is also of interest.
While there have been studies using wastewater data generated from infectious disease outbreaks in small populations, such as long term care facilities or college dormitories \citep{keck2024wastewater,acer2022quantifying}, to our knowledge, there are no statistical methodologies which address this challenge directly.
In this paper, we provide a novel method for efficiently estimating the time-varying effective reproduction number, the expected number of individuals a newly infected person would subsequently infect, in small populations from wastewater data. 
\par 
To account for the inherent uncertainty in the spread of infectious disease, we would ideally model an epidemic using a stochastic population model.
In practice, a Markov Jump Process (MJP) which counts the number of individuals in compartments representing susceptible, infectious and recovered individuals is a common choice \citep{andersson2012stochastic,allen2010introduction}. 
However, using MJP based models in inference tasks is computationally challenging when the state space is large and the process is only partially observed \citep{ho2018birth,rupp2024differentiated}.
In large populations, deterministic approximations, such as using a model described by ordinary differential equations (ODEs), or a model based on renewal equations, can be used, but in small populations, these deterministic approximations can be inadequate \citep{fintzi2022linear}. 
Most methods for estimating the effective reproduction number from wastewater data have relied on these deterministic approximations, and thus may perform poorly in small population settings. 
\par 
Exact inference of MJPs for epidemics is usually accomplished either through Sequential Monte Carlo or data augmentation coupled with Markov chain Monte Carlo \citep{o1999bayesian, fintzi2017efficient, wang2025bayesian, pomparticle,andrieu2010particle}. 
Both approaches can be computationally intensive and prone to implementation issues in practice, and improving these methodologies remains an active area of research \citep{corbella2022lifebelt,morsomme2022exact, wang2025bayesian}.
\par 
It is also common practice to approximate the MJP with the linear noise approximation (LNA), a local approximation to a stochastic differential equation whose solution is itself a diffusion process approximation of the MJP \citep{fearnhead2014inference,buckingham2018gaussian,fintzi2022linear,golightly2023accelerating}. 
The key advantage of the linear noise approximation is that it has Gaussian transition densities.
Another approach, proposed by \citet{isham1991assessing}, is to start with the Gaussian transition densities (justified by central limit theorem style arguments originating from the work of Kurtz \citep{kurtz1971limit,brittonandpardoux}), and then choose approximate first and second moments of the MJP to use as the moments of the Gaussian density \citep{isham1991assessing,buckingham2018gaussian}.
Both approaches are significantly more computationally intensive than deterministic approximations, as the latent epidemic states must now be inferred along with other parameters, but scale much better with population size than exact inference methods.
\par 
For the specific task of inferring the effective reproduction number, often an approximate model is chosen which does not model the number of susceptible individuals explicitly. 
The most popular methods are based on a branching process approximation to the MJP where individuals infect other individuals in an independent and identically distributed manner, giving rise to the commonly used renewal equation relating the current number of new cases to the previous counts of new cases \citep{cori_new_2013,epidemia_paper,goldstein2024incorporating}.
Avoiding modeling susceptibles simplifies the model significantly, while still allowing for accurate inference of the effective reproduction number.
Because these branching process based models are based on incidence, it can be cumbersome to connect them to data sources which are not explicitly realizations of incidence, such as wastewater data. 
Inspired by previous analyses using birth-death models used in phylodynamic analyses \citep{stadler2013birth,zhukova2023fast}, we previously implemented a deterministic compartmental model which modeled latent, infectious and recovered individuals but not susceptibles \citep{goldstein2024semiparametric}.
\par 
Here we adapt our previous approach by first defining an MJP without a susceptible compartment. 
This in turn greatly simplifies the use of either the LNA or the moment approximation technique, as the infinitesimal transition rates are now linear with respect to the states of the compartments of the MJP. 
In this paper, we chose to use the approach of \citet{isham1991assessing}, taking advantage of the simplified form of the MJP to calculate the exact conditional first and second moments in closed form. 
Our approach allows us to make use of state-of-the-art high dimensional Markov chain Monte Carlo methods which scale well with population size while still accounting for the stochastic nature of infectious disease transmission.
\par 
We compare our stochastic model to its deterministic counterpart, as well as a stochastic branching process-based model under multiple simulation scenarios. 
We first vary the shape of the curve we are trying to infer, then change the ``stochasticity" of the underlying epidemic dynamics by varying the initial number of infected individuals and the total population size.
Finally, we apply our wastewater based methods to estimate the effective reproduction number of SARS-CoV-2 in several college campus dormitories and compare the results with a state-of-the-art stochastic case-based method, demonstrating the qualitative differences between wastewater-based and case-based methods, as well as the differences between deterministic and stochastic methods in a small population setting. 
Overall, we find that analyzing wastewater data in small populations produces potentially useful insights into the infection dynamics of an epidemic, and that stochastic methods are clearly less biased and less prone to spuriously inferring dramatic changes in dynamics as compared to deterministic methods in the small population settings.

\section{Methods}
\subsection{Wastewater Data}
It is common practice to measure the concentration of pathogen genomes from the same sample of wastewater multiple times, producing multiple measurements called replicates. 
In publicly available databases and dashboards, a sample mean of replicates is often reported, but we will instead focus on the replicates themselves, as there are advantages to using the raw data as opposed to sample means \citep{goldstein2024semiparametric}. 
We define $\mathbf{X} = (X_{t_{1},1}, \dots, X_{t_{1},j}, \dots, X_{t_{T},j})$, where $X_{t_{i},k}$ is the $kth$ replicate of pathogen genomes collected from wastewater at time $t_{i}$, with units of RNA copies per milliliter. 
We will model $X_{t_{i},k}$ as a noisy realization of the unobserved number of currently infectious individuals.

\subsection{Stochastic Compartmental Models}
\subsection{The SEIR Model} \label {seirr_model}
The SEIR model describes an infectious disease outbreak of a homogeneously mixing population, with the population divided into four compartments: susceptible, exposed (infected but not yet infectious), infectious, and removed.
In its stochastic form, we represent the SEIR model as a four dimensional continuous time Markov jump process, $\mathbf{G(t)} = (S(t), E(t), I(t), R(t))$.
By construction, $R(t)$ is redundant, as $R(t) = N - S(t) - E(t) - I(t)$, where $N$ is the fixed total population size.
The SEIR dynamics can be defined in terms of rate parameters such that 
\begin{align*}
    P(\mathbf{G}(t + dt) &= (s - 1, e + 1, i, r)  \mid \mathbf{G}(t) = (s,e,i,r)) = \beta i s/N dt + o(dt),\\
    P(\mathbf{G}(t + dt) &= (s, e - 1, i + 1, r) \mid \mathbf{G}(t) = (s,e,i,r)) =  \gamma e  dt + o(dt), \\ 
    P(\mathbf{G}(t + dt) &= (s, e, i - 1, r + 1) \mid \mathbf{G}(t) = (s,e,i,r)) = \nu  i dt + o(dt),\\
    P(\mathbf{G}(t + dt) &= (s, e, i, r ) \mid \mathbf{G}(t) = (s,e,i,r)) = 1 - (\beta i  s/N  + \gamma  e  + \nu  i)dt + o(dt). 
\end{align*}
Here $\gamma$ is the inverse of the mean latent period, and $\nu$ is the inverse of the mean infectious period. 
We describe the infectiousness of the disease through the rate parameter $\beta$.
In practice, we will allow $\beta$ to be time-varying, and denote it $\beta_{t}$, to allow for changes in population or pathogen characteristics such as public health policies or emergence of more transmissible genetic variants.
With this model, the time-varying basic reproduction number, $R_{0,t}$, and effective reproduction number, $R_{t}$, are defined as
\begin{align*}
R_{0,t} = \frac{\beta_{t}}{\nu},  \quad R_{t} =  R_{0,t} \times \frac{S(t)}{N}.
\end{align*}
The basic reproduction number, $R_{0,t}$ is the expected number of individuals an individual infected at time $t$ would subsequently infect in a completely susceptible population. 
The effective reproduction number $R_{t}$ is the expected number of individuals an individual infected at time $t$ would subsequently infect if conditions remained the same as they were at time $t$. 
Intuitively, $R_{t}$ takes into account the fact that some people have already been infected, as it is $R_{0,t}$ multiplied by the fraction of the population which is still susceptible at time $t$.
\subsection{The EI Model}
We reduce the SEIR model to the EI model, and represent it as a two dimensional continuous time Markov jump process $\mathbf{H}(t) = (E(t), I(t))$, defined as:
\begin{align*}
    P(\mathbf{H}(t + dt) &= (e + 1, i)  \mid \mathbf{H}(t) = (e,i)) = \alpha_{t} i  dt + o(dt),\\
    P(\mathbf{H}(t + dt) &= (e - 1, i + 1) \mid \mathbf{H}(t) = (e,i)) =  \gamma  e dt + o(dt), \\ 
    P(\mathbf{H}(t + dt) &= (e, i - 1) \mid \mathbf{H}(t) = (e,i)) = \nu  i  dt + o(dt),\\
    P(\mathbf{H}(t + dt) &= (e, i) \mid \mathbf{H}(t) = (e,i)) = 1- (\alpha_{t}  i + \gamma  e  + \nu \times i) dt + o(dt).
\end{align*}
Note that $R_{t}$ is still recoverable by setting $\alpha_{t} = \beta_{t} \times \frac{S(t)}{N}$, so that
\begin{align*}
R_{t} =  R_{0,t} \times \frac{S(t)}{N}  = \frac{\alpha_{t}}{\nu}.
\end{align*}
This equality allows us to infer $R_{t}$ non-parametrically using the EI model. 
\par 
Unfortunately, the transition probabilities of $\mathbf{H}(t)$ are not analytically tractable. 
However, if we assume that $\alpha_{t}$ is piece-wise constant, then for any particular interval of time, the conditional moments of $\mathbf{H}(t)$ are available in closed form. 
We will use them to construct approximations to the transition probabilities of $\mathbf{H}(t)$, following the techniques of \citet{isham1991assessing}.
\subsection{Constructing a Partial Differential Equation of the Moment Generating Function}\label{moment_ode}

Let $p_{e,i}(t) = P(\mathbf{H}(t) = (e,i) | \mathbf{H}(0) = (x,y))$. 
We will omit indexing by $x,y$ for notational simplicity. 
Then the Kolmogorov Forward equation for $\mathbf{H}(t)$ is 
\begin{equation}\label{eqn:kmgf}
    \frac{dp_{e,i}(t)}{dt} = \alpha i p_{e-1,i}(t) + \gamma(e+1)p_{e+1,i-1}(t) + \nu(i+1)p_{e,i+1}(t) - (\alpha i + \gamma e + \nu i)p_{e,i}(t).
\end{equation}
From this differential equation with respect to time, we can construct a partial differential equation of the moment generating function of $\mathbf{H}(t)$, by multiplying both sides of Equation \ref{eqn:kmgf} by $e^{\theta_{1}e +\theta_{2}i}$, where $\theta_{1}, \theta_{2}$ take values in an interval of the real line which includes 0, and sum over all possible values of each of compartments. 
Let $M(\boldsymbol{\theta};t)$ be the Moment-generating function of $\mathbf{H}(t)$.
We produce the following partial differential equation: 
\begin{equation}
    \frac{dM(\boldsymbol{\theta};t)}{dt} = \left(\alpha e^{\theta_{1}}\frac{d}{d\theta_{2}}+\nu e^{-\theta_{2}}\frac{d}{d\theta_{2}} + \gamma e^{-\theta_{1}+ \theta_{2}}\frac{d}{d\theta_{1}} - \alpha \frac{d}{d\theta_{2}} - \nu \frac{d}{d\theta_{2}} -\gamma \frac{d}{d\theta_{1}}\right)M(\boldsymbol{\theta};t).
\end{equation}
A nearly identical procedure for constructing a partial differential equation of the probability generating function is shown in Chapter 9 of  \citet{renshaw2015stochastic} for the two-site birth death process in which the site-specific birth rate only depends on its own site (in contrast, in our model, the birth rate for the E compartment depends on the I compartment). 
The above equation can also be derived more directly through conditional expectations of the change in $\mathbf{H}(t)$ in infinitesimal intervals using Bailey's ``random variable trick" \citep{bailey1964elements}.
\par 
By taking the partial derivative on both sides, and setting $\boldsymbol{\theta} = \mathbf{0}$, we can create a system of linear ordinary differential equations for the conditional moments of the EI model (see Appendix section \ref{sec:ei_odes} for the series of ODEs).
We used Mathematica version 13.1 \citep{Mathematica} to generate closed form solutions of the conditional expectations, variances, and covariance.
Let $\boldsymbol{\mu}$ be the vector of conditional expectations, and $\boldsymbol{\Sigma}$ be the matrix of conditional variances and covariance.
We then use the derived conditional moments to construct densities which approximate the transition probabilities of the continuous time Markov jump process. 
\subsection{Log-Normal Approximation of the transition probabilities}
We start with the known result that when the compartment counts are large enough, the transition probability mass function converges to the normal density \citep{kurtz1971limit,barbour1974functional,brittonandpardoux}, that is, for $t > 0$
\begin{equation}
\mathbf{H}(t)\mid \mathbf{H}(0) \sim \text{Normal}(\boldsymbol{\mu}, \boldsymbol{\Sigma}).
\end{equation}
In practice, we wish to rule out the possibility of negative compartment counts. 
To do this, we instead use the transition density for the log compartment counts, and appeal to the delta method to construct the density. That is:
\begin{equation}
    \log{\mathbf{H}(t)}\mid \mathbf{H}(0) \sim \text{Normal}(\log{\boldsymbol{\mu}}, \mathbf{J}_{\log{\boldsymbol{\mu}}} \Sigma \mathbf{J}_{\log{\boldsymbol{\mu}}}),
\end{equation}
where $ \mathbf{J}_{\log{\boldsymbol{\mu}}}$ is the Jacobian of $\log{\boldsymbol{\mu}}$. 
% That is, if we we are interested in the transition density $P(\mathbf{H}(t)|\mathbf{H}(l))$ but $\alpha_{t}$ changes at time $j$ where $l < j < t$, we rewrite the desired density as 
% \begin{equation}
% P(\mathbf{H}(t)|\mathbf{H}(l)) = P(\mathbf{H}(t)|\mathbf{H}(j))\times P(\mathbf{H}(j)|\mathbf{H}(l)),
% \end{equation}
% which follows from the Markovian property.
\par
We place additional explicit priors on the latent state space for computational reasons. 
First, we require the compartment counts to be non-zero to avoid taking the log of 0. 
Second, we require the compartment counts and the means of the compartment counts to sum to less than 8 billion (approximate global human population) in order to avoid computations with infinity. 
\subsection{Observation Model}\label{eiww_obsmodel}
Recall that $X_{t_{i}, k}$ is the $k$th replicate of the concentration observed at time $t_{i}$. 
On the log scale, we model $X_{t_{i},k}$ as a noisy realization of the number of currently infectious individuals, where 
\begin{equation}
\log{X_{t_{i},k}} \sim \text{Normal}(\log{\left(I(t_{i})\right)} + \log{\left(\rho\right)}, \tau^{2}).
\end{equation}
Here $\rho$ is a scaling factor, $\tau$ is a noise parameter, both receive Log-Normal priors.

\subsection{Complete Stochastic EI-ww Model Structure}
We use a random walk prior for the time-varying effective reproduction number: $R_{0} \sim \text{Log-Normal}(\mu_{0}, \sigma_{0}), \sigma \sim \text{Log-Normal}(\mu_{rw}, \sigma_{rw}), \log{(R_{k_{i}})}|R_{k_{i-1}}, \sigma \sim \text{Normal}(\log{(R_{k_{i-1}})}, \sigma)$.
Let $\boldsymbol{\Theta} = (\gamma, \nu, I(0), E(0))$, $\mathbf{R} = (R_{k_{1}}, \dots, R_{k_{M}})$ be the vector of effective reproduction number values and $\mathbf{H} = (\mathbf{H}_{t_{i}}, \dots, \mathbf{H}_{t_{I}})$ be the matrix of latent states from $\mathbf{H}(t)$. 
We infer the states of $\mathbf{H}$ which correspond to times when we observe data and times at which $\alpha_{t}$ changes. 
The target posterior distribution is: 
\begin{equation*}
        P(\textbf{R},\mathbf{H}, \boldsymbol{\Theta}, \rho, \tau, \sigma \mid \textbf{X}) \propto
                \mathrlap{\underbrace{\phantom{P(\mathbf{X} \mid \mathbf{H}, \mathbf{R}, \boldsymbol{\Theta}, \rho, \tau)}}_{\text{Concentration Model}}}
P(\mathbf{X} \mid \mathbf{H}, \mathbf{R}, \boldsymbol{\Theta}, \rho, \tau)\mathrlap{\underbrace{\phantom{P(\mathbf{H} \mid \mathbf{R}, \boldsymbol{\Theta})}}_{\text{LN EI}}}P(\mathbf{H} \mid \mathbf{R}, \boldsymbol{\Theta})\mathrlap{\underbrace{\phantom{P(\mathbf{R} \mid \sigma)}}_{\text{RW Prior}}}P(\mathbf{R} \mid \sigma)P(\boldsymbol{\Theta}, \rho, \tau, \sigma).
\end{equation*}
We use the No-U-Turn Sampler, implemented in the \texttt{Julia} package \texttt{Turing} to approximate this posterior distribution \citep{NUTS,turing}.
We used non-centered re-parameterizations for all model parameters.
Markov chain Monte Carlo chains were initialized using the Maximum A Posterior (MAP) estimate of each parameter with added independent Gaussian noise.

\subsection{Other models}
We compare our stochastic EI-ww model to its deterministic counterpart. 
In the deterministic model, the compartments are modeled with a set of ODEs as follows
\begin{align*}
\frac{dE}{dt} &= \alpha I - \gamma E, \\
\frac{dI}{dt} &= \gamma E - \nu I. 
\end{align*}
These turn out to be the same as the ODEs for the first moments of $\mathbf{H}(t)$. 
Let $\mathbf{S}(\boldsymbol{\Theta}, t)$ be the solution to the ODEs. 
Then the posterior of interest for the deterministic EI-ww model is 
\begin{equation*}
 P(\textbf{R}, \boldsymbol{\Theta}, \rho,  \tau,  \sigma \mid \textbf{X}) \propto  P(\mathbf{X} \mid \mathbf{S}(\boldsymbol{\Theta}, t), \mathbf{R}, \boldsymbol{\Theta}, \rho, \tau)P(\mathbf{R} \mid \sigma)P(\boldsymbol{\Theta}, \rho, \tau, \sigma).
\end{equation*}
There is no additional term for the compartment counts, because they are deterministic functions of the other parameters.
\par 
As an additional comparison, we implement a branching process approximation model using the R package \texttt{Episewer} \citep{lison2024improving}.
While the package was originally developed for working with digital PCR data, it provides a number of different models which are appropriate for other types of data as well.
This model assumes pathogen concentrations are realizations of a convolution of unobserved latent incidence and a population level shedding load profile which describes how the mean individual's total shedding of genomes changes over the course of their infection. 
Most models in this class of methods assume that latent incidence evolves deterministically, but in our implementation, incidence is modeled as a conditionally normal random variable, with the mean-variance structure of the negative binomial distribution.
We will refer to this model as Episewer in the remainder of the text.
Episewer is described in more detail in Appendix Section \ref{sec:episewer}.
\par 
Finally, when analyzing real data, we analyze counts of new cases (case data) with a model constructed using the Epidemia package \citet{epidemia_paper}.
The model is an example of the common branching process based methods for estimating $R_{t}$ from cases, where the mean number of new infections is equal to a weighted sum of the previous new infections multiplied by the effective reproduction number.
Unlike most implementations of this class of methods, Epidemia allows latent incidence to be modeled stochastically, which we expect produces better inference in the small population setting.
The full model is written in Appendix Section \ref{sec:epidemia}, and more thorough descriptions of this class of models can be found in \citet{epidemia_paper} and \citet{goldstein2024incorporating}.
All code necessary to recreate the analyses of this paper are available at \url{https://github.com/igoldsteinh/not_flushed_away}.
\section{Simulations}
\subsection{Simulation Protocol}
We simulated data from an agent-based stochastic compartmental model with seven I compartments and 3 R compartments, which models the infection history of each individual in the population, but is equivalent to a population level SEIIIIIIIRRR model when aggregated (Appendix Section \ref{sec:sei7r}). 
All individuals in the seven I compartments and the first two R compartments shed pathogen RNA. 
The mean shedding within a compartment is constant. However, across compartments the mean changes according to biologically plausible trajectories (in brief: shedding peaks in the third I compartment, and rapidly decreases, with comparably little shedding occurring in the R compartments). In addition, each individual was assigned a random intercept to change their individual shedding trajectory from the mean in a consistent manner. 
The exact details are available in  Appendix Section \ref{sec:sim}, but the overall goal of our simulation method was to mimic what we believe are key characteristics of wastewater data: namely that shedding is not constant but changes over the course of an individual's infection, and that there is significant heterogeneity in shedding across individuals. 
The rates governing time spent in each compartment were chosen to mimic SARS-CoV-2 (Appendix Section \ref{sec:sim_params}). 
Wastewater concentrations were generated using the normal distribution as in Section \ref{eiww_obsmodel}, however the mean of the distribution was the total genome concentration of the population, calculated by aggregating the individual concentration shed by each individual in the population. 
\par 
For the first two simulations, the total population size was 1000 and we simulated epidemics under two different $R_{0,t}$ curves, one where $R_{0,t}$ changed from 1.1 to 2.1 in five weeks (Steep), and another where $R_{0,t}$ stayed fixed at 2 (Fixed). 
We initialized 20 individuals in the E compartment, with the remainder in the S compartment. 
For the Fixed $R_{0,t}$ scenario, we simulated an additional two scenarios; one where the total population size was 500, with an initial 10 in the E compartment, and one where the total population size was 2000 with an initial 40 in the E compartment (Total500 and Total2000 respectively). 
The latter two scenarios were meant to test how the models performed under varying levels of stochasticity. 
We would expect that smaller populations would have more stochastic variation, while the opposite would be true for larger populations. 
Models were fit to the first fourteen weeks of data, data was collected every other day, and we used three wastewater samples per day as data.
For all scenarios, for our models, we model $R_{t}$ as changing on a weekly basis. 
For each scenario, we simulated 100 epidemics. 
MCMC algorithms were initially run for 1000 iterations. 
If chains failed to pass basic convergence diagnostics, the models were re-fit using a larger number of iterations. 
For three simulations for the stochastic EI-ww model, only three chains were used, as the fourth failed to converge. 
An example simulation from the shallow curve setting is shown in Appendix Figure \ref{fig:sim_data}.
\subsection{Simulation Results}
\begin{figure}[H]
    \centering
    \includegraphics[width=\textwidth]{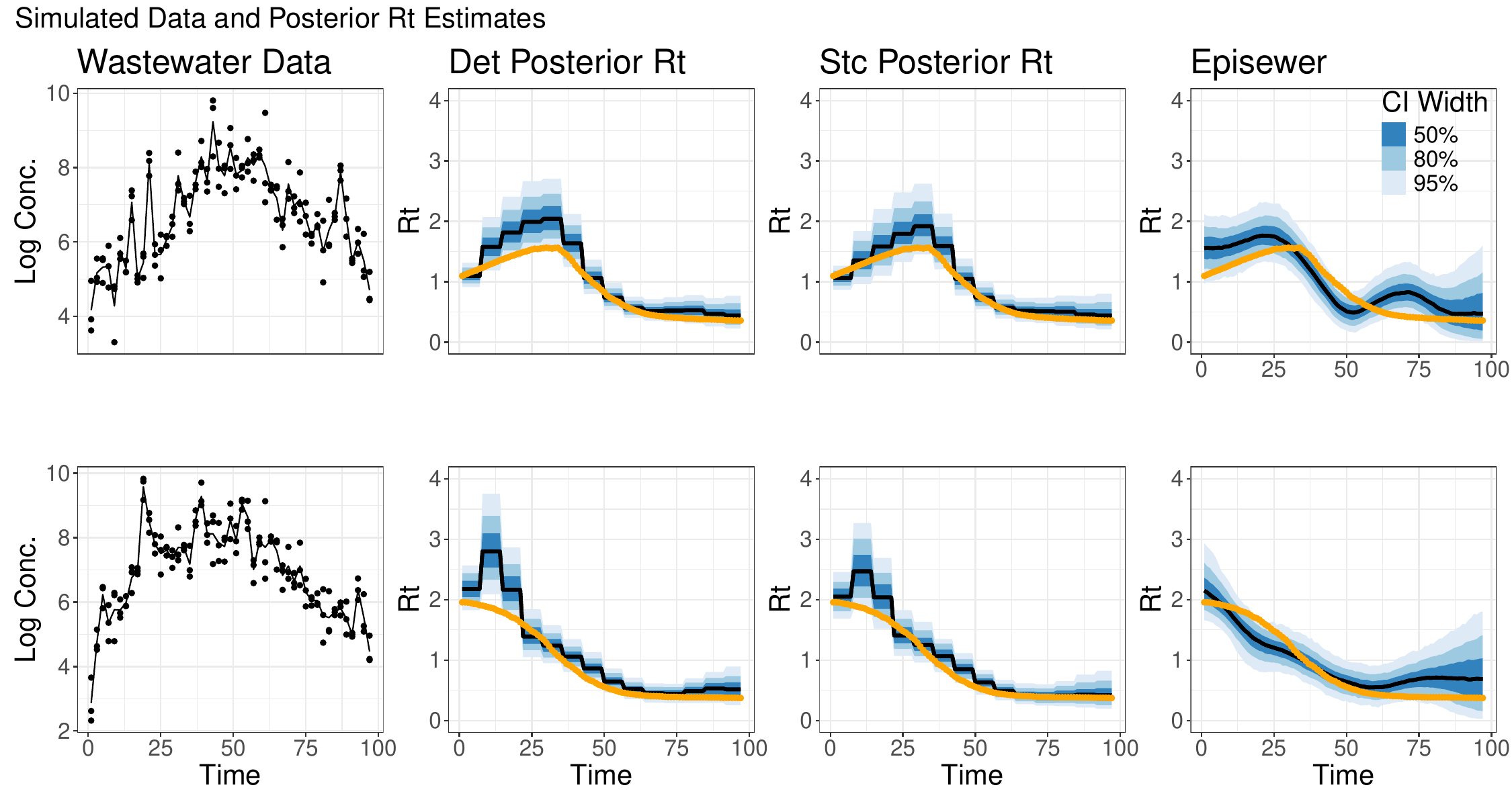}
    \caption{Posterior summaries of $R_{t}$ using the stochastic and deterministic EI-ww models for the Steep and Fixed $R_{t}$ curves. 
    The first column is the simulated data models were fit to. 
    The second column shows the posterior summaries for the deterministic EI-ww model, the third column shows the posterior summaries of the stochastic EI-ww model, and the fourth column shows the posterior summaries of the Episewer model. 
    First row is the Steep true $R_{t}$ curve, second row is the Fixed true $R_{t}$ curve. 
    True $R_{t}$ values are shown in orange, black lines are posterior medians, blue shaded areas are credible intervals.}
    \label{fig:sample_rt}
\end{figure}

Example posterior medians and credible intervals from the deterministic and stochastic EI-ww models, as well as from Episewer for the Steep and Fixed scenarios are shown in Figure \ref{fig:sample_rt}.
In both scenarios, the deterministic EI-ww model is worse at covering the true shape of the $R_{t}$ curve, inferring overly dramatic changes in $R_{t}$. 
Episewer generally has wider credible intervals than the stochastic EI-ww model, and follows the curve less closely than the stochastic EI-ww model. 
Episewer appears to perform better in the Fixed scenario as opposed to the Steep scenario. 
\par
\begin{figure}[H]
    \centering
    \includegraphics[width=\textwidth]{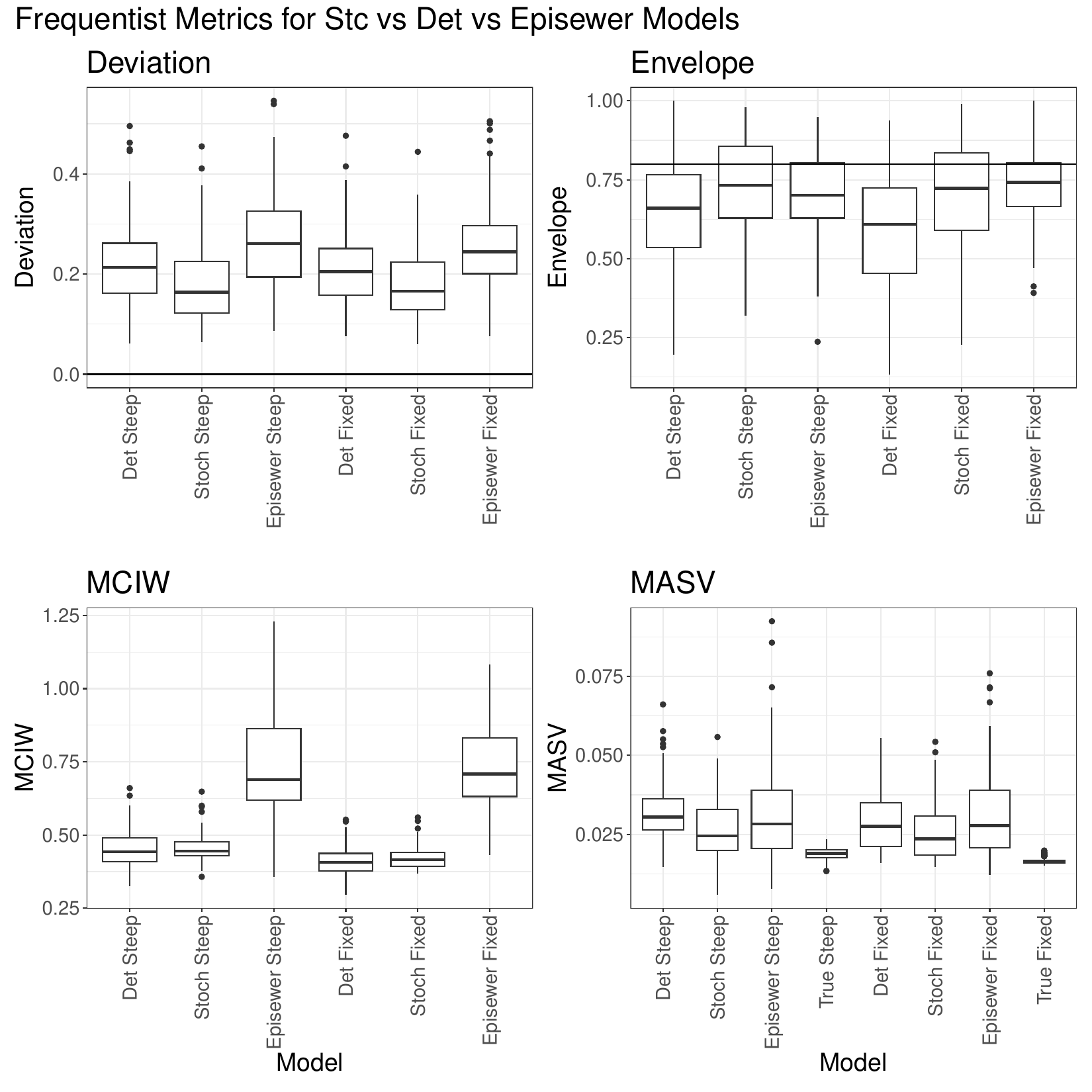}
    \caption{Frequentist metrics for the deterministic, stochastic, and Episewer models in the Steep and Fixed curve scenarios. 
     For the top row and bottom left plots, the x-axis describes the model and scenario, for the bottom right plot (MASV), the x-axis includes extra values for the true MASV for each simulation scenario.
     Absolute deviation is the mean of the absolute value of the difference between the median $R_{t}$ at each time point and the true value. Envelope is a measure of coverage, taking the mean coverage of 80\% intervals over the time series. MCIW is the mean of the mean credible interval width. Mean absolute standard deviation (MASV) is the difference between the current median point estimate for $R_{t}$ and the previous point estimate for $R_{t}$, we compare it against the true MASV for each simulation.}
    \label{fig:freq_baseline}
\end{figure}

 
We summarise the performance of our models across all 100 data sets for each scenario by reporting some frequentist metrics, shown in Figure \ref{fig:freq_baseline}. 
For each simulation the envelope is a measure of coverage, and is the proportion of time points for which an 80\% credible interval from the posterior distribution captured the true value of interest. 
Mean credible interval width (MCIW) is the mean of 80\% credible interval widths across time points within a simulation. 
Absolute deviation is a measure of accuracy, and is the mean of the absolute difference between the posterior median and the true value at each time point. 
Finally, mean absolute sequential variation (MASV) measures the variation in the effective reproduction number across time by computing the mean of the absolute difference between the posterior median of $R_{t}$ at $t$ and the posterior median at $t-1$. 
Each simulation has its own true MASV, which is the difference between the true $R_{t}$ at $t$ and $t-1$, we summarise the true MASV with its own box-plot, the closer the box-plot of the model's MASV matches the box-plot of the true MASV, the better.
In both scenarios, the stochastic model is more accurate than either the deterministic or Episewer models, however it is more uncertain than the deterministic model. 
Episewer is the most uncertain of all three models. 
While the stochastic model is better calibrated than the deterministic model, its median Envelope is a little below 0.8 in both scenarios.
In the case of the Fixed scenario, Episewer is slightly better calibrated, however in the Steep scenario the stochastic EI-ww model is slightly better calibrated.
The stochastic EI-ww model is less prone to sudden changes than either the deterministic or Episewer models, and thus its MASV is closer to the true MASV in both scenarios. 
The stochastic EI-ww model performs reasonably well across both scenarios, while the deterministic model is clearly worse.
While Episewer is consistently more biased than the stochastic EI-ww model, it is sometimes better calibrated, though at the cost of wider credible intervals.
\par
\begin{figure}[H]
    \centering
\includegraphics[width=\textwidth]{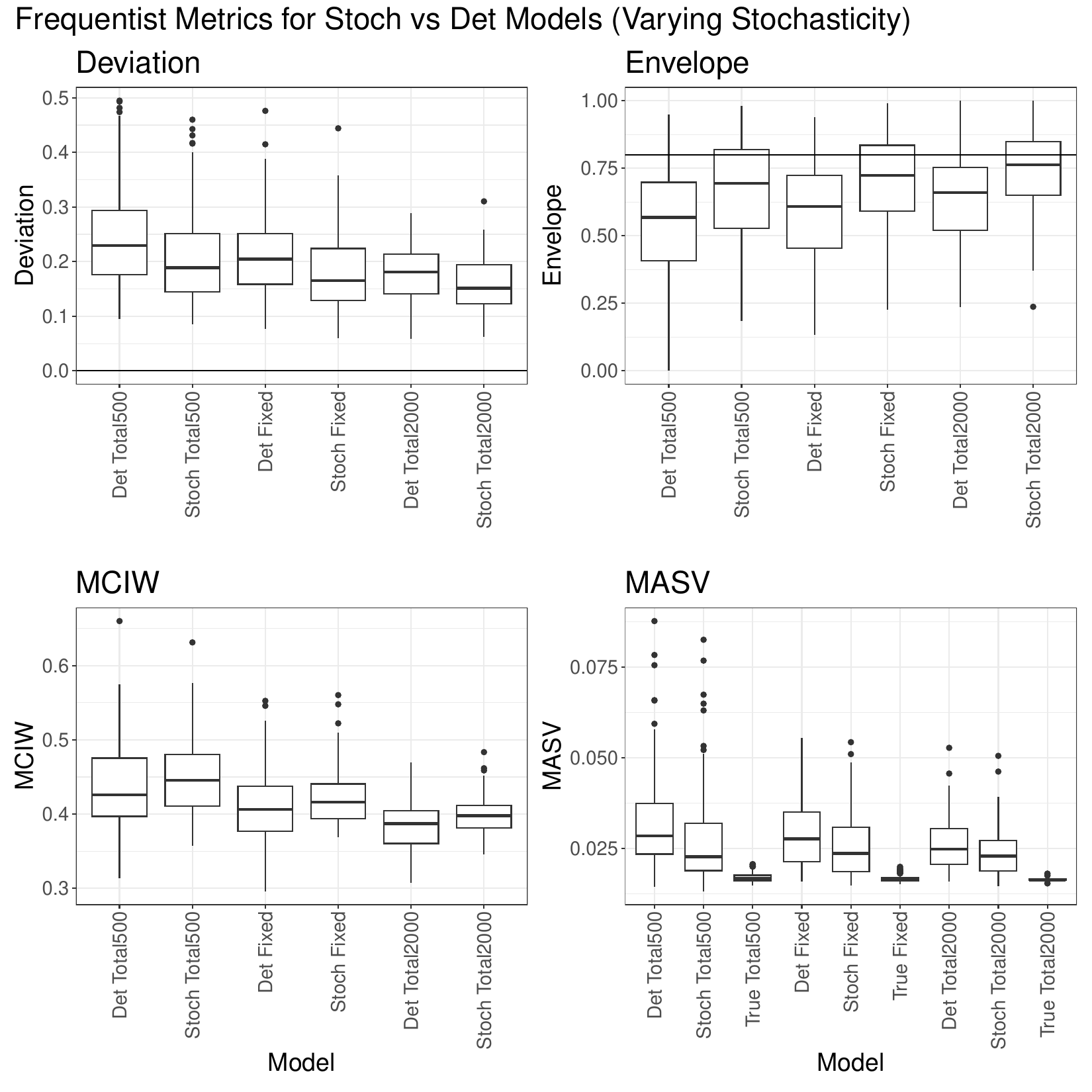}
    \caption{Frequentist metrics for the deterministic and stochastic models with varying initial and total populations for the Fixed scenario.
     For the top row and bottom left plots, the x-axis describes the model and scenario, for the bottom right plot (MASV), the x-axis includes extra values for the true MASV for each simulation scenario.
     All scenarios are variations on the Fixed scenario with $R_{0,t} = 2$. 
     Total500 and Total2000 refer to the total population size (500 and 2000 respectively), as well as the initial number E individuals (10 and 40 respectively). 
     The original Fixed scenario with 20 individuals in the E and total population size 1000 is included for reference. 
    See Figure \ref{fig:freq_baseline} for metric definitions.}
    \label{fig:freq_stoch}
\end{figure}

 
We wanted to explore how stochasticity in the model affected the differences in performance. 
To this end, we kept the shape of the $R_{t}$ curve the same but changed the stochasticity of the epidemic by varying the total number of individuals in the population, as well as the number of individuals starting in the E compartment.
For this comparison, we chose to focus on just the stochastic and deterministic EI-ww models. 
These differences are summarised in Figure \ref{fig:freq_stoch}. 
The general model differences seen in previous scenarios (more accuracy, more uncertainty) remain consistent. 
In general, the stochastic model is more uncertain and more accurate than the deterministic model. 
\section{The Effective Reproduction Number in UC Irvine Residential Communities}
Surveillance data from the SARS-CoV-2 pandemic at the University of California, Irvine were available between Dec 2021 and June 2022 \citep{pappu2025tracking,uci_ww_data, pappu2025wastewater}.
About 860 wastewater samples were collected from 13 different student housing communities on the University of California Irvine campus from December 2021 to June 2022. These samples were analyzed for SARS-CoV-2 N2 and E genes and water quality parameters such as TSS, COD and ammonia. 
Usually three replicates were tested per day, if sample analysis of any of the three replicates failed, all three replicates were excluded from data analysis.
In addition, counts of new cases were also available.
While wastewater data were available at a sub-community level (e.g. individual buildings or spatial regions of a community), case data were only available at the community level.
\par 
We chose to analyze two sub-communities and one community with populations around 1000. 
The total size of the G community is around 2400, which includes sub-communities G1 and G2. The total size of the E community is around 1090, which is served by a single sewer manhole. 
The G community and the E community are on opposite sides of the campus, both house undergraduate students. 
While data from December and January were available, we chose not to analyze these data as the campus was on winter break, and then delayed in person classes for the first few weeks of the winter quarter, and so the campus residential population was not stable. 
We also expect that many reported cases were from individuals returning from travel and testing positive, thus not representing local transmission events.
We chose to not analyze data from June 2022 as there were no case data available for June.
UCI had a policy of randomly testing individuals in the population, which was discontinued in mid-March of 2022. 
This change in testing policy is accounted for in our case model by modeling the case detection rate as dependent on an indicator variable which equals 0 before the change in policy, and 1 after the change in policy. 
\par
\begin{figure}[H]
    \centering
    \includegraphics[width = 1.0\textwidth]{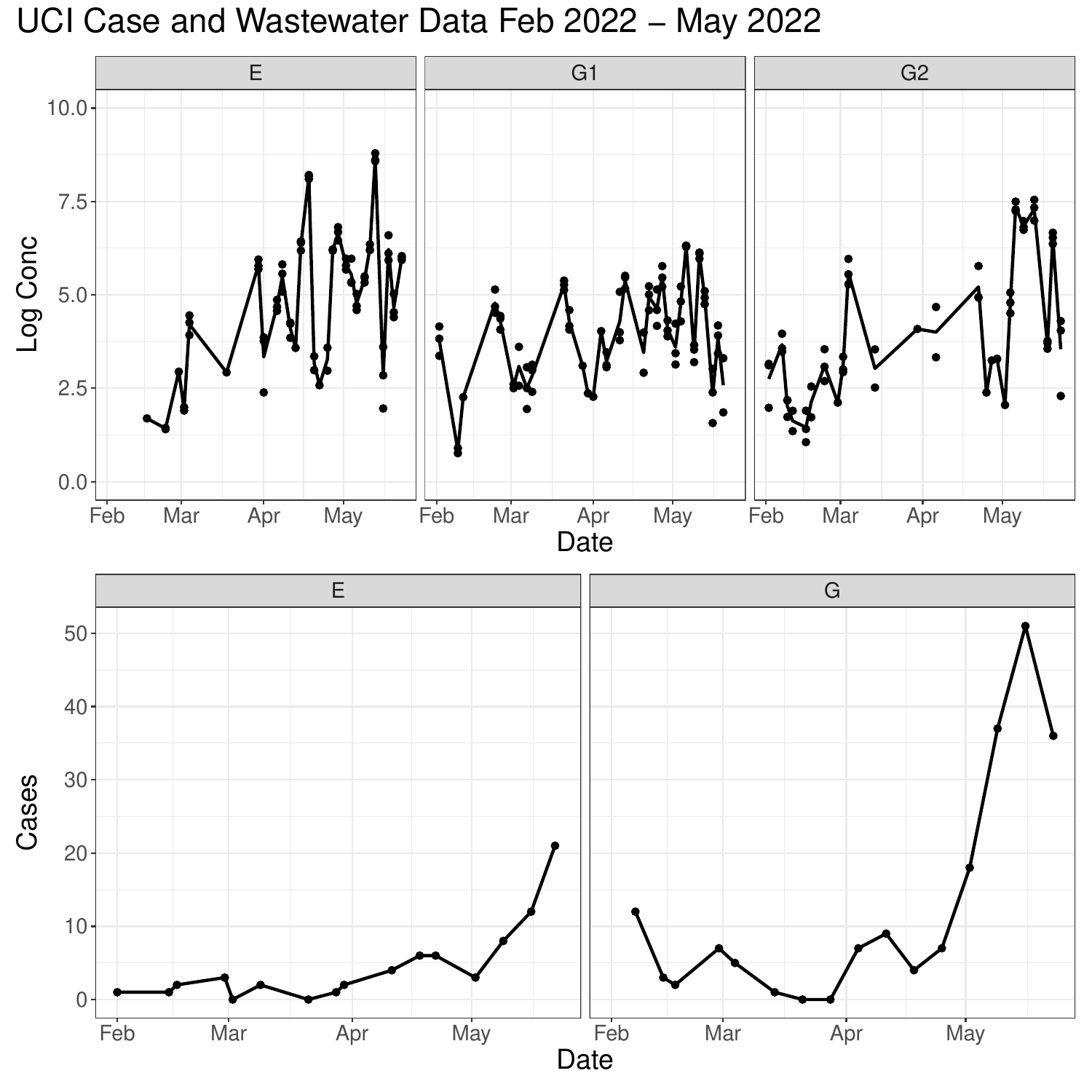}
    \caption{Log concentrations of SARS-CoV-2 RNA and weekly reported new COVID-19 cases at UC Irvine for February 2022 through May 2022. For the log concentrations, the dots are individual replicates, and the lines connect the means.
    The cases are reported at the community level, while the concentrations are reported at the sub-community level for community G. G1 and G2 are sub-communities within the larger G community, E is a separate community with no sub-communities.}
    \label{fig:uci_data}
\end{figure}
\begin{figure}[H]
    \centering
    \includegraphics[width = 1.0\textwidth]{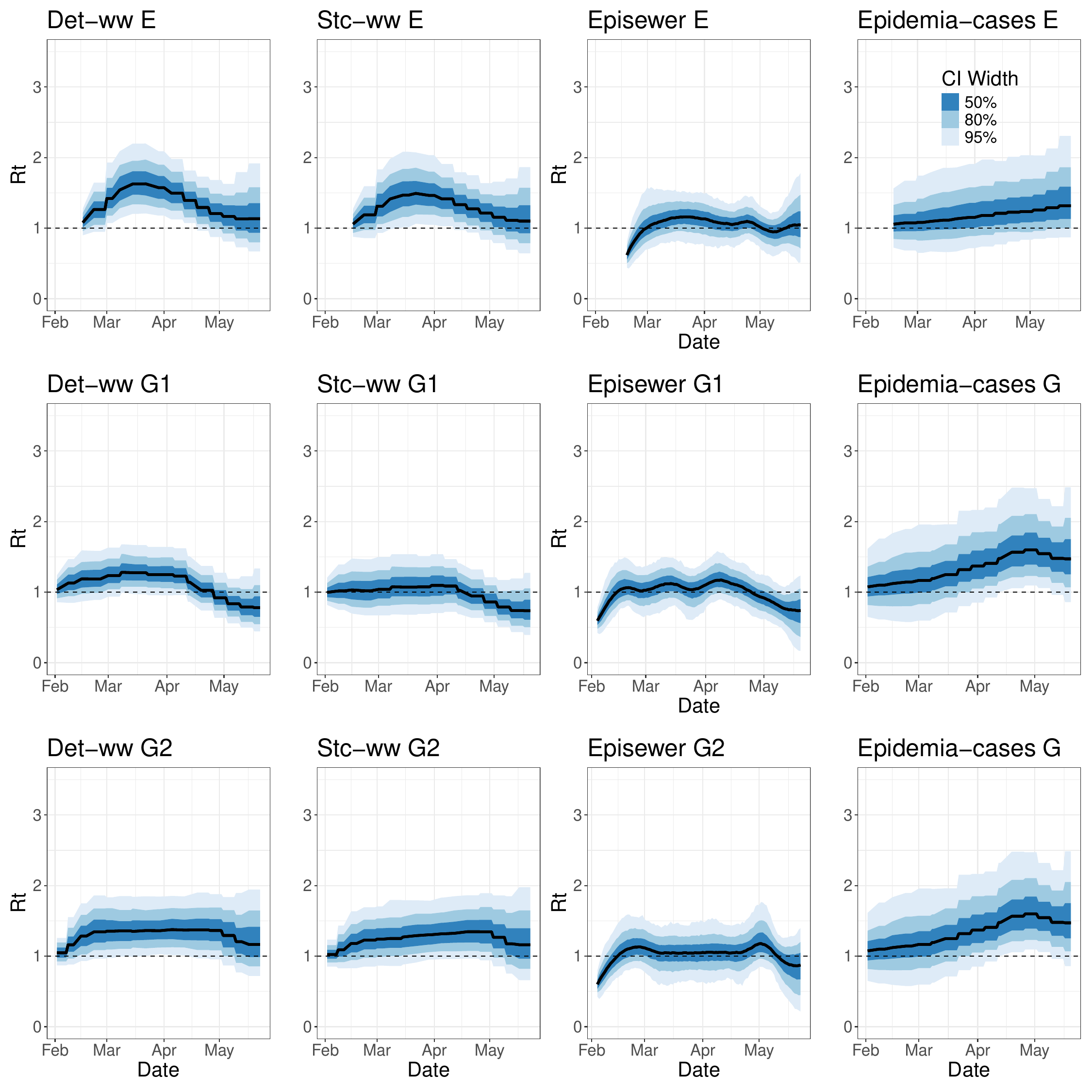}
    \caption{Posterior summaries of the effective reproduction number in college campus communities estimated from wastewater data only (stochastic, deterministic, and Episewer models) or case data only (Epidemia-cases). 
    Black lines are posterior medians, blue shaded regions are credible intervals. 
    G1 and G2 are sub-populations of the G campus community, the E campus community has no sub-communities.
    The bottom two panels in the fourth column are identical.}
    \label{fig:uci_posterior_rt}
\end{figure}

We fit the stochastic and deterministic EI-ww wastewater models, the Episewer Model, and the Epidemia case model to these data sets. 
The priors for the EI-ww models were the same as in simulations, except for the initial conditions and initial $R_{t}$ which was centered around the posterior median of $R_{t}$ for Orange County, CA (where UC Irvine is located) from a previous analysis using case and test data \citep{goldstein2024incorporating}.
Likewise, the priors for Episewer were the same except for the prior on initial $R_{t}$.
Prior specifications for all models are available in Appendix Sections \ref{sec:sim_priors} and \ref{sec:episewer_priors}.
The data are displayed in Figure \ref{fig:uci_data}, and the posterior trajectories of $R_{t}$ are summarised in Figure \ref{fig:uci_posterior_rt}.

The deterministic model shows more dramatic changes in $R_{t}$ than either of the stochastic models.
There is also clear variability in the estimated $R_{t}$ between different communities and sub-communities, for example, looking at the EI-ww models, we might reasonably conclude that G2 experienced an outbreak between March and May, while it would be hard to conclude the same about G1. 
These differences are averaged over in our case-based estimate of the whole community. 
The EI-ww models seem to detect an increase in $R_{t}$ before the case-based models do in G2 and E1. 
The posterior medians of the wastewater models are above 1 weeks or months before the posterior medians of the case-based models.
The stochastic EI-ww wastewater-based posterior estimates are still uncertain enough that the 95\% credible intervals are never above 1. 
\par 
Comparing Episewer to to the stochastic EI-ww model, we find the two models generally disagree, with the most substantial disagreements in the G1 and G2 communities. 
The stochastic EI-ww model estimates steady increases and decreases, whereas the Episewer model estimates shorter and more frequent increases and decreases in $R_{t}$. 
The Episewer and Epidemia-cases model report very different results in all three sub-communities.
\section{Discussion}
%we made a stochastic model to work in small population settings
%it seems to be a little more uncertain and less biased than deterministic version
%it is certainly less volatile
%the small population setting is difficult, the range of coverage for both models was quite large
% this is likely due to model misspecification
%There are qualitative differences between case and wastewater models in real data settings
%uncertainty is still quite high, but at the very least, a public health officer comparing the two approaches might want to order some exploratory tests
%speaks to the need for models which can combine information to help reduce uncertainty
%I think one takeaway is that pandemics happen at fairly fine spatial resolution, and its worth doing what we can to provide fine grained analyses (maybe compare this to the OC Rt estimate in this time frame)
%still a need for understanding nowcasting performance
In this paper we developed a model with a stochastic latent epidemic process and compared it against the deterministic version and a branching process inspired model in small population settings. 
The stochastic EI-ww model was generally more accurate and more uncertain than the deterministic EI-ww model, leading to overall better frequentist calibration. 
The stochastic EI-ww model was also generally more accurate and less uncertain than the Episewer model, though depending on the shape of the underlying $R_{t}$ curve, this lead to slightly better or worse overall calibration.
We also used our models to estimate the effective reproduction number of SARS-CoV-2 at UC Irvine, showing qualitatively different estimates between all three models, as well as differences between models using wastewater versus case data. 
\par 
The stochastic EI-ww model performed generally better than the deterministic EI-ww model. 
The wide range of envelope values also speaks to highly variable (and in some cases quite poor) performance on individual realizations of the different simulations. 
We speculate this wide variability in performance arises from the inherently noisy data, and the fact that our model is misspecified, i.e. it does not take into account the time-varying dynamics of pathogen genome shedding \citep{hoffmann2021faecal}.
We expect this issue of misspecification to be more acute in small populations, where a change in state for a single individual can lead to large variation in overall shedding.
\par
Our model was at least competitive, and sometimes clearly better  than, the Episewer model which does explicitly account for the time-varying nature of shedding through the use of a shedding load profile, suggesting mitigating some of this model misspecification could indeed improve performance.
In terms of the noisiness of the data, obtaining narrower posterior credible intervals may require using multiple data sources, models which make use of spatial correlation to combine information, or models which are less misspecified than our current approach  (or a combination of all three). 
% \par 
% For our Episewer model, we used an observation model with no explicit mean variance relationship, as our simulated data did not have such a relationship. 
% The package documentation strongly discourages the use of this observation model, instead advocating for using an explicit mean variance relationship.
% Since our simulations did not use an explicit mean variance relationship, we found this led to much worse model performance (analysis not shown).
% We recommend caution when specifying mean-variance relationships unless one has high confidence there is not much model misspecification in other parts of the model. 
\par 
In our real data application, while $R_{t}$ estimates were often uncertain when using wastewater data, they were still qualitatively different than the estimates obtained from case data. 
The 95\% credible intervals from the stochastic EI-ww model always went below 1 throughout the observation period, on the other hand, in the case of E and G2, the curve estimated from wastewater data shows with reasonably high probability $R_{t}$ was above 1 weeks before the case model. 
These results suggest using wastewater data in small populations has the promise to provide actionable insights not obtainable from using case data alone.
\par 
The deterministic EI-ww model provided similar results to the stochastic EI-ww model, although it inferred steeper increases in $R_{t}$ and occasionally had credible intervals above 1. 
The Episewer model, on the other hand, generally did not infer clear increases or decreases in $R_{t}$ during the modeling period. 
Given our experiments, we view the stochastic EI-ww model results as more conservative and more accurate than the deterministic EI-ww model, and less conservative but more accurate than the Episewer model. 
\par 
In addition, our study demonstrates that epidemic dynamics vary based on community and sub-communities. 
Surveillance at different levels of population aggregation reveal different dynamics, which in turn require different actions from public health officials. 
Developing methodologies that continue to improve performance at high spatial resolution seems like a fruitful area of future research. 
\par 
In our model, we required that the compartment counts be non-zero. 
This limits our model to situations where we plausibly believe the pathogen is circulating among a non-zero number of individuals. 
Developing an alternative method which does not require this restriction is an important adaptation of our approach to pursue.

\bibliographystyle{abbrvnat}
%\vspace{-0.9cm}
\section*{Acknowledgments}
%This was was supported in part by......
ARP and SJ were supported by UCI COVID CARFT and EPA-G2021-STAR-A1, Grant number: 84025701. 
DMP was supported by funding through the Gates Foundation (INV-028123) and the National Institutes of Health (FAIN: U19AI089672).
IHG was supported by a Stanford Center for Computational, Evolutionary and Human Genomics Fellowship. 
VMM was supported by the National Institutes of Health grant R01AI170204.
This work utilized the resources of the Research Cyberinfrastructure Center (RCIC) at UC Irvine.
We thank Jamie Jones for useful discussions that, among other things, inspired the title of this paper.

\bibliography{../../references}

\begin{thebibliography}{57}
\providecommand{\natexlab}[1]{#1}
\providecommand{\url}[1]{\texttt{#1}}
\expandafter\ifx\csname urlstyle\endcsname\relax
  \providecommand{\doi}[1]{doi: #1}\else
  \providecommand{\doi}{doi: \begingroup \urlstyle{rm}\Url}\fi

\bibitem[Acer et~al.(2022)Acer, Kelly, Lover, and Butler]{acer2022quantifying}
P.~T. Acer, L.~M. Kelly, A.~A. Lover, and C.~S. Butler.
\newblock {Quantifying the relationship between {SARS-CoV-2} wastewater
  concentrations and building-level {COVID-19} prevalence at an isolation
  residence: a passive sampling approach}.
\newblock \emph{International Journal of Environmental Research and Public
  Health}, 19\penalty0 (18):\penalty0 11245, 2022.

\bibitem[Allen(2010)]{allen2010introduction}
L.~J. Allen.
\newblock \emph{{An Introduction to Stochastic Processes with Applications to
  Biology}}.
\newblock {CRC press}, 2010.

\bibitem[Andersson and Britton(2012)]{andersson2012stochastic}
H.~Andersson and T.~Britton.
\newblock \emph{{Stochastic Epidemic Models and their Statistical Analysis}},
  volume 151.
\newblock Springer Science \& Business Media, 2012.

\bibitem[Andrieu et~al.(2010)Andrieu, Doucet, and
  Holenstein]{andrieu2010particle}
C.~Andrieu, A.~Doucet, and R.~Holenstein.
\newblock Particle {M}arkov chain {M}onte {C}arlo methods.
\newblock \emph{Journal of the Royal Statistical Society Series B: Statistical
  Methodology}, 72\penalty0 (3):\penalty0 269--342, 2010.

\bibitem[Bailey(1964)]{bailey1964elements}
N.~T. Bailey.
\newblock \emph{The elements of stochastic processes with applications to the
  natural sciences}.
\newblock John Wiley \& Sons, 1964.

\bibitem[Barbour(1974)]{barbour1974functional}
A.~D. Barbour.
\newblock On a functional central limit theorem for {M}arkov population
  processes.
\newblock \emph{Advances in Applied Probability}, 6\penalty0 (1):\penalty0
  21--39, 1974.

\bibitem[Bhatt et~al.(2023)Bhatt, Ferguson, Flaxman, Gandy, Mishra, and
  Scott]{epidemia_paper}
S.~Bhatt, N.~Ferguson, S.~Flaxman, A.~Gandy, S.~Mishra, and J.~A. Scott.
\newblock {Semi-Mechanistic {B}ayesian modeling of {COVID}-19 with Renewal
  Processes}.
\newblock \emph{Journal of the Royal Statistical Society Series A: Statistics
  in Society}, 186\penalty0 (4):\penalty0 601--615, 2023.

\bibitem[Britton and Pardoux(2019)]{brittonandpardoux}
T.~Britton and E.~Pardoux.
\newblock Stochastic epidemics in a homogenous community.
\newblock In T.~Britton and E.~Pardoux, editors, \emph{Stochastic Epidemic
  Models with Inference}, Lecture Notes in Mathematics (LNM), volume 2255,
  pages 3--119. Springer, 2019.

\bibitem[Buckingham-Jeffery et~al.(2018)Buckingham-Jeffery, Isham, and
  House]{buckingham2018gaussian}
E.~Buckingham-Jeffery, V.~Isham, and T.~House.
\newblock Gaussian process approximations for fast inference from infectious
  disease data.
\newblock \emph{Mathematical {B}iosciences}, 301:\penalty0 111--120, 2018.

\bibitem[Champredon and Dushoff(2015)]{champredon2015}
D.~Champredon and J.~Dushoff.
\newblock Intrinsic and realized generation intervals in infectious-disease
  transmission.
\newblock \emph{Proceedings of the Royal Society B: Biological Sciences},
  282\penalty0 (1821):\penalty0 2015--2026, 2015.

\bibitem[Champredon et~al.(2018)Champredon, Dushoff, and Earn]{champredon2018}
D.~Champredon, J.~Dushoff, and D.~J.~D. Earn.
\newblock Equivalence of the {E}rlang-distributed {SEIR} epidemic model and the
  renewal equation.
\newblock \emph{SIAM Journal on Applied Mathematics}, 78\penalty0 (6):\penalty0
  3258--3278, 2018.

\bibitem[Corbella et~al.(2022)Corbella, McKinley, Birrell, Presanis, Spencer,
  Roberts, and De~Angelis]{corbella2022lifebelt}
A.~Corbella, T.~J. McKinley, P.~J. Birrell, A.~M. Presanis, S.~E. Spencer,
  G.~O. Roberts, and D.~De~Angelis.
\newblock The lifebelt particle filter for robust estimation from low-valued
  count data.
\newblock \emph{arXiv preprint arXiv:2212.04400}, 2022.

\bibitem[Cori et~al.(2013)Cori, Ferguson, Fraser, and Cauchemez]{cori_new_2013}
A.~Cori, N.~M. Ferguson, C.~Fraser, and S.~Cauchemez.
\newblock A new framework and software to estimate time-varying reproduction
  numbers during epidemics.
\newblock \emph{American Journal of Epidemiology}, 178\penalty0 (9):\penalty0
  1505--1512, 2013.

\bibitem[Fearnhead et~al.(2014)Fearnhead, Giagos, and
  Sherlock]{fearnhead2014inference}
P.~Fearnhead, V.~Giagos, and C.~Sherlock.
\newblock Inference for reaction networks using the linear noise approximation.
\newblock \emph{Biometrics}, 70\penalty0 (2):\penalty0 457--466, 2014.

\bibitem[Fintzi et~al.(2017)Fintzi, Cui, Wakefield, and
  Minin]{fintzi2017efficient}
J.~Fintzi, X.~Cui, J.~Wakefield, and V.~N. Minin.
\newblock Efficient data augmentation for fitting stochastic epidemic models to
  prevalence data.
\newblock \emph{{Journal of Computational and Graphical Statistics}},
  26\penalty0 (4):\penalty0 918--929, 2017.

\bibitem[Fintzi et~al.(2022)Fintzi, Wakefield, and Minin]{fintzi2022linear}
J.~Fintzi, J.~Wakefield, and V.~N. Minin.
\newblock A linear noise approximation for stochastic epidemic models fit to
  partially observed incidence counts.
\newblock \emph{Biometrics}, 78\penalty0 (4):\penalty0 1530--1541, 2022.

\bibitem[Ge et~al.(2018)Ge, Xu, and Ghahramani]{turing}
H.~Ge, K.~Xu, and Z.~Ghahramani.
\newblock Turing: A language for flexible probabilistic inference.
\newblock In A.~Storkey and F.~Perez-Cruz, editors, \emph{Proceedings of the
  Twenty-First International Conference on Artificial Intelligence and
  Statistics}, volume~84 of \emph{Proceedings of Machine Learning Research},
  pages 1682--1690. PMLR, 09--11 Apr 2018.

\bibitem[Gillespie(1977)]{gillespie1977exact}
D.~T. Gillespie.
\newblock Exact stochastic simulation of coupled chemical reactions.
\newblock \emph{{The Journal of Physical Chemistry}}, 81\penalty0
  (25):\penalty0 2340--2361, 1977.

\bibitem[Goldstein et~al.(2024{\natexlab{a}})Goldstein, Parker, Jiang, and
  Minin]{goldstein2024semiparametric}
I.~H. Goldstein, D.~M. Parker, S.~Jiang, and V.~M. Minin.
\newblock Semiparametric inference of effective reproduction number dynamics
  from wastewater pathogen surveillance data.
\newblock \emph{Biometrics}, 80\penalty0 (3):\penalty0 ujae074,
  2024{\natexlab{a}}.

\bibitem[Goldstein et~al.(2024{\natexlab{b}})Goldstein, Wakefield, and
  Minin]{goldstein2024incorporating}
I.~H. Goldstein, J.~Wakefield, and V.~M. Minin.
\newblock Incorporating testing volume into estimation of effective
  reproduction number dynamics.
\newblock \emph{Journal of the Royal Statistical Society Series A: Statistics
  in Society}, 187\penalty0 (2):\penalty0 436--453, 2024{\natexlab{b}}.

\bibitem[Golightly et~al.(2023)Golightly, Wadkin, Whitaker, Baggaley, Parker,
  and Kypraios]{golightly2023accelerating}
A.~Golightly, L.~E. Wadkin, S.~A. Whitaker, A.~W. Baggaley, N.~G. Parker, and
  T.~Kypraios.
\newblock Accelerating {B}ayesian inference for stochastic epidemic models
  using incidence data.
\newblock \emph{Statistics and Computing}, 33\penalty0 (6):\penalty0 134, 2023.

\bibitem[Han et~al.(2020)Han, Seong, Kim, Shin, Im~Cho, Park, Kim, Park, and
  Choi]{han2020viral}
M.~S. Han, M.-W. Seong, N.~Kim, S.~Shin, S.~Im~Cho, H.~Park, T.~S. Kim, S.~S.
  Park, and E.~H. Choi.
\newblock Viral {RNA} load in mildly symptomatic and asymptomatic children with
  {COVID-19, Seoul, South Korea}.
\newblock \emph{Emerging {I}nfectious {D}iseases}, 26\penalty0 (10):\penalty0
  2497, 2020.

\bibitem[Hillary et~al.(2020)Hillary, Malham, McDonald, and
  Jones]{hillary2020wastewater}
L.~S. Hillary, S.~K. Malham, J.~E. McDonald, and D.~L. Jones.
\newblock {Wastewater and public health: the potential of wastewater
  surveillance for monitoring COVID-19}.
\newblock \emph{Current Opinion in Environmental Science \& Health},
  17:\penalty0 14--20, 2020.

\bibitem[Ho et~al.(2018)Ho, Xu, Crawford, Minin, and Suchard]{ho2018birth}
L.~S.~T. Ho, J.~Xu, F.~W. Crawford, V.~N. Minin, and M.~A. Suchard.
\newblock Birth/birth-death processes and their computable transition
  probabilities with biological applications.
\newblock \emph{Journal of {M}athematical {B}iology}, 76:\penalty0 911--944,
  2018.

\bibitem[Hoffman and Gelman(2014)]{NUTS}
M.~D. Hoffman and A.~Gelman.
\newblock The {N}o-{U}-{T}urn sampler: Adaptively setting path lengths in
  {H}amiltonian {M}onte {C}arlo.
\newblock \emph{Journal of Machine Learning Research}, 15\penalty0
  (47):\penalty0 1593--1623, 2014.

\bibitem[Hoffmann and Alsing(2023)]{hoffmann2021faecal}
T.~Hoffmann and J.~Alsing.
\newblock Faecal shedding models for {SARS-CoV-2 RNA} among hospitalised
  patients and implications for wastewater-based epidemiology.
\newblock \emph{Journal of the Royal Statistical Society Series C: Applied
  Statistics}, 72\penalty0 (2):\penalty0 330--345, 2023.

\bibitem[Isham(1991)]{isham1991assessing}
V.~Isham.
\newblock Assessing the variability of stochastic epidemics.
\newblock \emph{Mathematical {B}iosciences}, 107\penalty0 (2):\penalty0
  209--224, 1991.

\bibitem[Jin et~al.(2024)Jin, Tay, Ng, Wong, and Cook]{jin2024combining}
S.~Jin, M.~Tay, L.~C. Ng, J.~C.~C. Wong, and A.~R. Cook.
\newblock Combining wastewater surveillance and case data in estimating the
  time-varying effective reproduction number.
\newblock \emph{Science of The Total Environment}, 928:\penalty0 172469, 2024.

\bibitem[Keck et~al.(2024)Keck, Adatorwovor, Liversedge, Mijotavich, Olsson,
  Strike, Amirsoleimani, Noble, Torabi, Rockward, et~al.]{keck2024wastewater}
J.~W. Keck, R.~Adatorwovor, M.~Liversedge, B.~Mijotavich, C.~Olsson, W.~D.
  Strike, A.~Amirsoleimani, A.~Noble, S.~Torabi, A.~Rockward, et~al.
\newblock Wastewater surveillance for identifying sars-cov-2 infections in
  long-term care facilities, kentucky, usa, 2021--2022.
\newblock \emph{Emerging Infectious Diseases}, 30\penalty0 (3):\penalty0 530,
  2024.

\bibitem[Killingley et~al.(2022)Killingley, Mann, Kalinova, Boyers,
  Goonawardane, Zhou, Lindsell, Hare, Brown, Frise,
  et~al.]{killingley2022safety}
B.~Killingley, A.~J. Mann, M.~Kalinova, A.~Boyers, N.~Goonawardane, J.~Zhou,
  K.~Lindsell, S.~S. Hare, J.~Brown, R.~Frise, et~al.
\newblock Safety, tolerability and viral kinetics during {SARS-CoV-2} human
  challenge in young adults.
\newblock \emph{Nature Medicine}, 28\penalty0 (5):\penalty0 1031--1041, 2022.

\bibitem[King et~al.(2016)King, Nguyen, and Ionides]{pomparticle}
A.~A. King, D.~Nguyen, and E.~L. Ionides.
\newblock Statistical inference for partially observed {Markov} processes via
  the {R} package {pomp}.
\newblock \emph{Journal of Statistical Software}, 69\penalty0 (12):\penalty0
  1--43, 2016.
\newblock \doi{10.18637/jss.v069.i12}.

\bibitem[Kurtz(1971)]{kurtz1971limit}
T.~G. Kurtz.
\newblock Limit theorems for sequences of jump {M}arkov processes approximating
  ordinary differential processes.
\newblock \emph{Journal of Applied Probability}, 8\penalty0 (2):\penalty0
  344--356, 1971.

\bibitem[Lison et~al.(2024)Lison, Julian, and Stadler]{lison2024improving}
A.~Lison, T.~R. Julian, and T.~Stadler.
\newblock Improving inference in wastewater-based epidemiology by modelling the
  statistical features of digital {PCR}.
\newblock \emph{bioRxiv}, pages 2024--10, 2024.

\bibitem[Lui et~al.(2020)Lui, Ling, Lai, Tso, Fung, Chan, Ho, Luk, Chen, Ng,
  et~al.]{lui2020viral}
G.~Lui, L.~Ling, C.~K. Lai, E.~Y. Tso, K.~S. Fung, V.~Chan, T.~H. Ho, F.~Luk,
  Z.~Chen, J.~K. Ng, et~al.
\newblock Viral dynamics of {SARS-CoV-2} across a spectrum of disease severity
  in {COVID-19}.
\newblock \emph{Journal of Infection}, 81\penalty0 (2):\penalty0 318--356,
  2020.

\bibitem[Miyazawa et~al.(2024)Miyazawa, Wong, Ito, Iwamoto, Watanabe, van
  Boven, Wallinga, and Miura]{miyazawa2024wastewater}
S.~Miyazawa, T.~S. Wong, G.~Ito, R.~Iwamoto, K.~Watanabe, M.~van Boven,
  J.~Wallinga, and F.~Miura.
\newblock Wastewater-based reproduction numbers and projections of {COVID}-19
  cases in three areas in {J}apan, {N}ovember 2021 to {D}ecember 2022.
\newblock \emph{Eurosurveillance}, 29\penalty0 (8):\penalty0 2300277, 2024.

\bibitem[Morsomme and Xu(2022)]{morsomme2022exact}
R.~Morsomme and J.~Xu.
\newblock Exact inference for stochastic epidemic models via uniformly ergodic
  block sampling.
\newblock \emph{arXiv preprint arXiv:2201.09722}, 2022.

\bibitem[Morvan et~al.(2022)Morvan, Jacomo, Souque, Wade, Hoffmann, Pouwels,
  Lilley, Singer, Porter, Evens, et~al.]{morvan2022analysis}
M.~Morvan, A.~L. Jacomo, C.~Souque, M.~J. Wade, T.~Hoffmann, K.~Pouwels,
  C.~Lilley, A.~C. Singer, J.~Porter, N.~P. Evens, et~al.
\newblock An analysis of 45 large-scale wastewater sites in {England} to
  estimate {SARS-CoV-2} community prevalence.
\newblock \emph{Nature Communications}, 13\penalty0 (1):\penalty0 4313, 2022.

\bibitem[Nourbakhsh et~al.(2022)Nourbakhsh, Fazil, Li, Mangat, Peterson,
  Daigle, Langner, Shurgold, D’Aoust, Delatolla,
  et~al.]{nourbakhsh2022wastewater}
S.~Nourbakhsh, A.~Fazil, M.~Li, C.~S. Mangat, S.~W. Peterson, J.~Daigle,
  S.~Langner, J.~Shurgold, P.~D’Aoust, R.~Delatolla, et~al.
\newblock {A wastewater-based epidemic model for SARS-CoV-2 with application to
  three Canadian cities}.
\newblock \emph{Epidemics}, 39:\penalty0 100560, 2022.

\bibitem[Okita et~al.(2022)Okita, Morita, and Kumanogoh]{okita2022duration}
Y.~Okita, T.~Morita, and A.~Kumanogoh.
\newblock Duration of {SARS-CoV-2 RNA} positivity from various specimens and
  clinical characteristics in patients with {COVID}-19: a systematic review and
  meta-analysis.
\newblock \emph{Inflammation and Regeneration}, 42\penalty0 (1):\penalty0
  1--19, 2022.

\bibitem[O’Neill and Roberts(1999)]{o1999bayesian}
P.~D. O’Neill and G.~O. Roberts.
\newblock Bayesian inference for partially observed stochastic epidemics.
\newblock \emph{Journal of the Royal Statistical Society Series A: Statistics
  in Society}, 162\penalty0 (1):\penalty0 121--129, 1999.

\bibitem[Pappu et~al.(2025{\natexlab{a}})Pappu, Green, Oakes, and
  Jiang]{pappu2025tracking}
A.~R. Pappu, A.~Green, M.~Oakes, and S.~Jiang.
\newblock Tracking {COVID-19} trends in communities with low population by
  wastewater-based surveillance.
\newblock \emph{Science of The Total Environment}, 970:\penalty0 179007,
  2025{\natexlab{a}}.

\bibitem[Pappu et~al.(2025{\natexlab{b}})Pappu, Green, Oakes, and
  Jiang]{pappu2025wastewater}
A.~R. Pappu, A.~Green, M.~Oakes, and S.~Jiang.
\newblock {W}astewater-based surveillance data to determine the {COVID}-19
  trends in communities with low population.
\newblock \emph{Data in Brief}, page 111756, 2025{\natexlab{b}}.

\bibitem[Pappu et~al.(2025{\natexlab{c}})Pappu, Green, Oakes, and
  Jiang]{uci_ww_data}
A.~R. Pappu, A.~Green, M.~Oakes, and S.~Jiang.
\newblock {UCI} wastewater-based surveillance data ({SARS-CoV-2} and {PMMoV}
  datasets), 2025{\natexlab{c}}.
\newblock URL \url{https://data.mendeley.com/datasets/6x5zffpxwb/3}.

\bibitem[Polo et~al.(2020)Polo, Quintela-Baluja, Corbishley, Jones, Singer,
  Graham, and Romalde]{polo2020making}
D.~Polo, M.~Quintela-Baluja, A.~Corbishley, D.~L. Jones, A.~C. Singer, D.~W.
  Graham, and J.~L. Romalde.
\newblock Making waves: {W}astewater-based epidemiology for
  {COVID}-19--approaches and challenges for surveillance and prediction.
\newblock \emph{Water {R}esearch}, 186:\penalty0 116404, 2020.

\bibitem[Renshaw(2015)]{renshaw2015stochastic}
E.~Renshaw.
\newblock \emph{{Stochastic Population Processes: Analysis, Approximations,
  Simulations}}.
\newblock {OUP Oxford}, 2015.

\bibitem[Rupp et~al.(2024)Rupp, Schill, S{\"u}skind, Georg, Klever, L{\"o}sch,
  Grasedyck, Wettig, and Spang]{rupp2024differentiated}
K.~Rupp, R.~Schill, J.~S{\"u}skind, P.~Georg, M.~Klever, A.~L{\"o}sch,
  L.~Grasedyck, T.~Wettig, and R.~Spang.
\newblock Differentiated uniformization: A new method for inferring {M}arkov
  chains on combinatorial state spaces including stochastic epidemic models.
\newblock \emph{Computational Statistics}, pages 1--21, 2024.

\bibitem[Sender et~al.(2022)Sender, Bar-On, Park, Noor, Dushoff, and
  Milo]{sender2022unmitigated}
R.~Sender, Y.~Bar-On, S.~W. Park, E.~Noor, J.~Dushoff, and R.~Milo.
\newblock The unmitigated profile of {COVID}-19 infectiousness.
\newblock \emph{Elife}, 11:\penalty0 e79134, 2022.

\bibitem[Song et~al.(2021)Song, Reinke, Hoxsey, Jackson, Krikorian, Melitas,
  Rosso, and Jiang]{song2021detection}
Z.~Song, R.~Reinke, M.~Hoxsey, J.~Jackson, E.~Krikorian, N.~Melitas, D.~Rosso,
  and S.~Jiang.
\newblock Detection of {SARS-CoV-2} in wastewater: Community variability,
  temporal dynamics, and genotype diversity.
\newblock \emph{American Chemical Society Environmental Science and Techonology
  Water}, 1\penalty0 (8):\penalty0 1816--1825, 2021.

\bibitem[Stadler et~al.(2013)Stadler, K{\"u}hnert, Bonhoeffer, and
  Drummond]{stadler2013birth}
T.~Stadler, D.~K{\"u}hnert, S.~Bonhoeffer, and A.~J. Drummond.
\newblock Birth–death skyline plot reveals temporal changes of epidemic
  spread in {HIV} and hepatitis {C virus (HCV)}.
\newblock \emph{Proceedings of the National Academy of Sciences}, 110\penalty0
  (1):\penalty0 228--233, 2013.

\bibitem[Svensson(2007)]{Svensson2007}
A.~Svensson.
\newblock A note on generation times in epidemic models.
\newblock \emph{Mathematical {Biosciences}}, 208\penalty0 (1):\penalty0
  300--311, 2007.

\bibitem[Walsh et~al.(2020)Walsh, Jordan, Clyne, Rohde, Drummond, Byrne, Ahern,
  Carty, O'Brien, O'Murchu, et~al.]{walsh2020sars}
K.~A. Walsh, K.~Jordan, B.~Clyne, D.~Rohde, L.~Drummond, P.~Byrne, S.~Ahern,
  P.~G. Carty, K.~K. O'Brien, E.~O'Murchu, et~al.
\newblock {SARS-CoV-2} detection, viral load and infectivity over the course of
  an infection.
\newblock \emph{Journal of Infection}, 81\penalty0 (3):\penalty0 357--371,
  2020.

\bibitem[Wang and Walker(2025)]{wang2025bayesian}
S.~Wang and S.~G. Walker.
\newblock Bayesian data augmentation for partially observed stochastic
  compartmental models.
\newblock \emph{Bayesian Analysis}, 20\penalty0 (1):\penalty0 1409--1432, 2025.

\bibitem[W{\"o}lfel et~al.(2020)W{\"o}lfel, Corman, Guggemos, Seilmaier, Zange,
  M{\"u}ller, Niemeyer, Jones, Vollmar, Rothe, et~al.]{wolfel2020virological}
R.~W{\"o}lfel, V.~M. Corman, W.~Guggemos, M.~Seilmaier, S.~Zange, M.~A.
  M{\"u}ller, D.~Niemeyer, T.~C. Jones, P.~Vollmar, C.~Rothe, et~al.
\newblock Virological assessment of hospitalized patients with {COVID-2019}.
\newblock \emph{Nature}, 581\penalty0 (7809):\penalty0 465--469, 2020.

\bibitem[Wolfram~Research()]{Mathematica}
I.~Wolfram~Research.
\newblock Mathematica, {V}ersion 13.1.
\newblock URL \url{https://www.wolfram.com/mathematica}.
\newblock Champaign, IL, 2022.

\bibitem[Xin et~al.(2022)Xin, Li, Wu, Li, Lau, Qin, Wang, Cowling, Tsang, and
  Li]{Xin2021}
H.~Xin, Y.~Li, P.~Wu, Z.~Li, E.~H. Lau, Y.~Qin, L.~Wang, B.~J. Cowling, T.~K.
  Tsang, and Z.~Li.
\newblock Estimating the latent period of coronavirus disease 2019
  {(COVID-19)}.
\newblock \emph{Clinical Infectious Diseases}, 74\penalty0 (9):\penalty0
  1678--1681, 2022.

\bibitem[Zhang et~al.(2021)Zhang, Cen, Hu, Du, Hu, Kim, and
  Dai]{zhang2021prevalence}
Y.~Zhang, M.~Cen, M.~Hu, L.~Du, W.~Hu, J.~J. Kim, and N.~Dai.
\newblock Prevalence and persistent shedding of fecal {SARS-CoV-2 RNA} in
  patients with {COVID-19} infection: A systematic review and meta-analysis.
\newblock \emph{{Clinical and Translational Gastroenterology}}, 12\penalty0
  (4), 2021.

\bibitem[Zhukova et~al.(2023)Zhukova, Hecht, Maday, and
  Gascuel]{zhukova2023fast}
A.~Zhukova, F.~Hecht, Y.~Maday, and O.~Gascuel.
\newblock Fast and accurate maximum-likelihood estimation of multi-type
  birth--death epidemiological models from phylogenetic trees.
\newblock \emph{Systematic Biology}, 72\penalty0 (6):\penalty0 1387--1402,
  2023.

\end{thebibliography}

\clearpage

\appendix

\setcounter{table}{0}
\setcounter{equation}{0}
\setcounter{section}{0}
\setcounter{figure}{0}

\renewcommand\thefigure{\thesection\-\arabic{figure}}
\renewcommand\thetable{\thesection\-\arabic{table}}

\section{Appendix}
\subsection{EI Moment ODEs}\label{sec:ei_odes}
We suppress the conditional nature of the moments, as well as the dependence on $t$.
\begin{align*}
\frac{dE[E]}{dt} &= \alpha E[I] - \gamma E[E],\\
\frac{dE[I]}{dt} &= \gamma E[E] - \nu E[I],\\
\frac{d[E^{2}]}{dt} &= \alpha E[I] + \gamma E[E] + 2 \alpha E[EI] - 2\gamma E[E^{2}], \\
\frac{dE[I^{2}]}{dt} &= \nu E[I] - 2\nu E[I^{2}] + \gamma E[E] + 2\gamma E[EI],\\
\frac{dE[EI]}{dt} &= \alpha E[I^{2}] - \nu E[EI] - \gamma E[E] + \gamma E[E^{2}] - \gamma E[EI]. \\
\end{align*}
Using the definition of variance and covariance, we arrive at ODEs for the conditional variance and covariance as well:
\begin{align*}
\frac{d\text{Var}(E)}{dt} &= \alpha E[I] - 2\gamma \text{Var}(E) + \gamma E[E] +  2\alpha \text{Cov}(E,I),\\
\frac{d\text{Var}(I)}{dt} &= \nu E[I] - 2\nu \text{Var}(I) + \gamma E[E] + 2\gamma \text{Cov}(E,I), \\
\frac{d\text{Cov}(E,I)}{dt} &= \alpha \text{Var}(I) + \gamma \text{Var}(E) - \gamma E[E] - (\nu + \gamma) \text{Cov}(E,I).
\end{align*}
\subsection{Simulation Comparison of MJP vs Log-Normal}
To visualize the difference between our Log-Normal process and the true Markov jump process, we simulated 1000 data sets from each model using a fixed value of $R_{t} = 1.5$ with 10, 20, and 30 individuals starting in the E and I compartments. 
We simulated from the MJP using the classic Gillespie algorithm \citep{gillespie1977exact}.
Marginal quantiles of the counts in each compartment at times 1 through 30 are displayed in Appendix Figure \ref{fig:mjp_vs_LN}.
\begin{figure}[H]
    \centering
    \includegraphics[width=\textwidth]{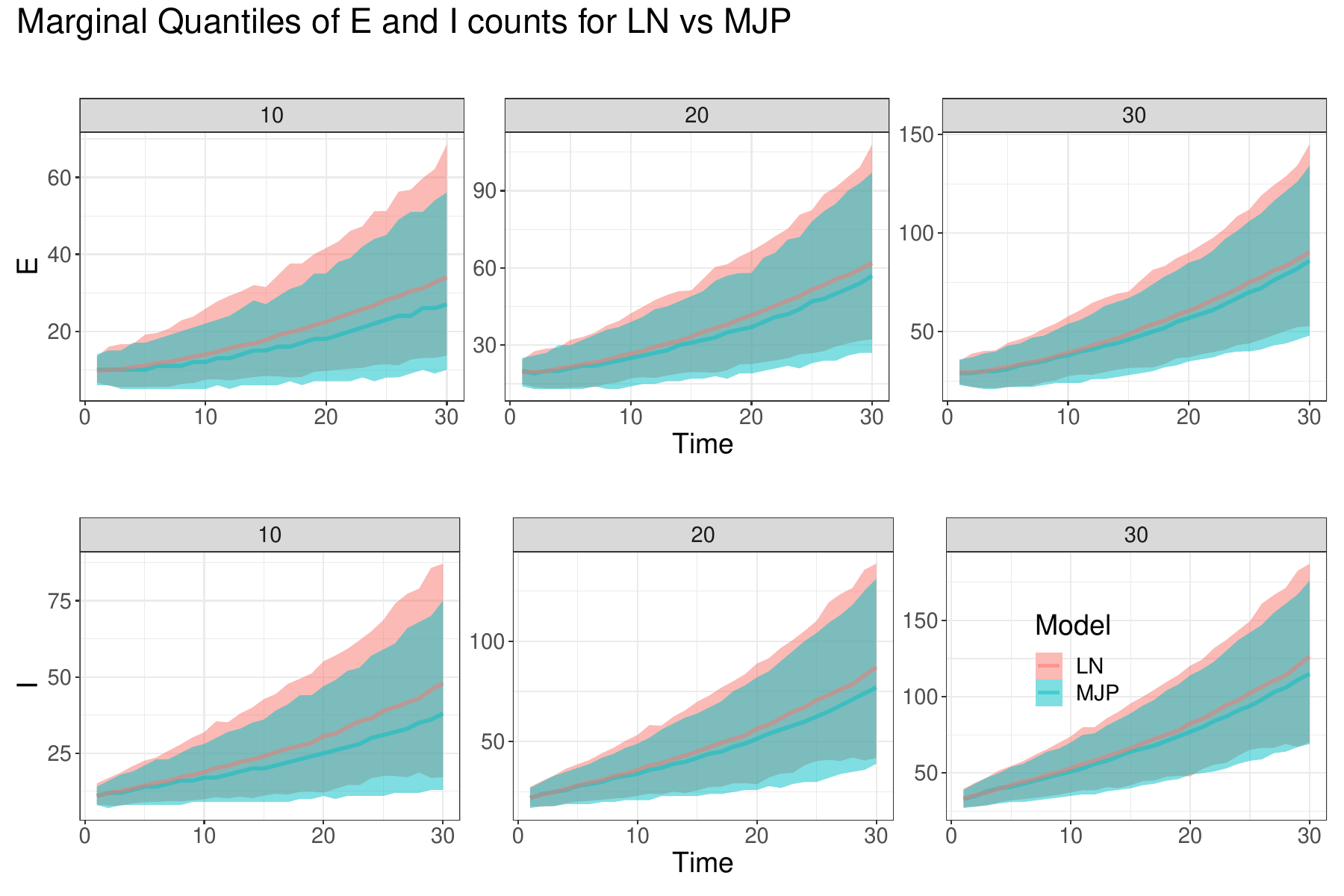}
    \caption{Empirical marginal quantiles of the counts in the E and I compartments for the original MJP EI model and the LN approximate model simulated from 1000 data sets. 
    The lines are medians, the ends of the shaded regions are the 2.5\% and 97.5\% quantiles.
    Each panel shows the distribution with a different number of individuals in the E and I compartments at the start of the simulation, all other parameters in the models are identical.
    As the number of individuals increases, the two distributions look more similar.}
    \label{fig:mjp_vs_LN}
\end{figure}
As we would expect, as the initial number of individuals in the simulation increases, the distributions of the MJP and Log-Normal approximate model look increasingly similar \citep{barbour1974functional}. 
\subsection{Obvservation Models using t-Distribution versus Normal Distribution}
We chose to use a Normal distribution as opposed to a t distribution to model observed concentrations, as we had previously done in \citet{goldstein2024semiparametric}. 
The t-distribution has fatter tails than the normal distribution, however the stochastic model has a latent stochastic epidemic process, which should accommodate additional levels of noise as well. 
Additionally, we expect that in small populations, the wastewater data may actually be less prone to outliers than in large populations, simply because the data is collected much closer to the source in small population settings. 
That is, we expect less degradation and transformation due to time spent in the sewer system, as the journey from toilet to sampler is shorter in a small population setting. 
As additional evidence that the normal distribution is adequate, we include the posterior predictive summaries of our models fit to real world data, which show no signs that the model is unable to simulate real data (Figure \ref{fig:uci_postpred}).

\subsection{Epidemia-cases}\label{sec:epidemia}
The Epidemia-cases model relies on the so-called renewal equation which calculates current incidence as a product of a weighted sum of previous incidence and the effective reproduction number. 
Let $I_{t}$ be the incidence at time $t$, $R_{t}$ be the effective reproduction number at time $t$, and $g(t)$ be the probability density function of the generation time distribution (the time between an individual becoming infected and infecting another individual; under the compartmental model framework this is usually taken to be equivalent to the sum of the latent period and the infectious period \citep{Svensson2007, champredon2015, champredon2018}). 
Then the classic renewal equation is:
\begin{equation*}
    E[I_{t}|I_{1}, \dots, I_{t-1}] = R_{t}\sum_{s=1}^{t-1}I_{s}g(t-s).
\end{equation*}
The \texttt{epidemia} package can be used to create different branching process inspired models to estimate the effective reproduction number using different observation models and models for latent incidence \citep{epidemia_paper}.
For the model we used in this study, we modeled observed cases using a negative binomial distribution, modeled the effective reproduction number as a Gaussian random walk, and modeled unobserved incidence as an auto-regressive normal random variable with variance equal to the mean multiplied by an over-dispersion parameter.
In additional, we modeled the case detection rate as a change point model with a baseline value plus an indicator function with a covariate, reflecting the change in UCI testing policy which occurred in March 2022.
The explicit model is listed below:
 \begin{align*}
\tau &\sim \text{exp}(\lambda)  \text{--Hyperprior for unobserved incidence,}\\
I_{\nu} &\sim \text{exp}(\tau) \text{--Prior on unobserved incidence $\nu$ days before observation,}\\
I_{\nu+1}, \dots, I_{0} &= I_{\nu}  \text{--Unobserved incidence,}\\
    \sigma & \sim \text{Truncated-Normal}(0, 0.15)\text{--Prior on variance of random walk} \\
   \log{R_{0}} &\sim \text{Normal}(\log{0.5}, 0.1)\text{--Prior on $R_{0}$,} \\
    \log{R_{t}}|\log{R_{t-1}} &\sim \text{Normal}(\log{R_{t-1}}, \sigma) \text{--Random walk prior on $R_{t}$,}\\
\psi &\sim \text{Normal}(10,2) \text{--Prior on variance parameter for incidence,} \\
I_{t}|I_{\nu}, \dots, I_{t-1}, R_{t} &\sim \text{Normal}(R_{t}\sum_{s<t}I_{s}g_{t-s}, \psi) \text{--Model for incidence,} \\
\text{logit} \alpha &= \beta_{0} + \beta_{1}\text{Policy Change} \text{--Case detection rate model,} \\
\beta_{0} & \sim \text{Normal}(0, 0.2) \text{--Baseline case detection rate,} \\
\beta_{1} & \sim \text{Normal}(0, 0.5) \text{--Difference in case detection post policy change,}\\
y_{t} &= \alpha \sum_{s<t}I_{s}\pi_{t-s} \text{--Mean of observed data model,}\\
\phi & \sim P(\phi) \text{--Prior on dispersion parameter for observed data,} \\
Y_{t} &\sim \text{Neg-Binom}(y_{t}, \phi) \text{--Observed data model.}\\
\end{align*}
Here $\pi_{t}$ are the values of the probability density function for the delay distribution, the time between an individual being infected and being observed.
\subsection{Episewer}\label{sec:episewer}
$\iota$ is the mean incidence. 
There is a random walk on the mean incidence before observation. 
Then the familiar renewal equation once observation begins 
\begin{equation*}
\iota_{t} = E[I_{t}] = R_{t} \sum_{s = 1}^{G} \tau_{s}^{gen}I_{t-s}|t>G.
\end{equation*}
Where $G$ is the maximum generation time.
Both models for incidence are normal models, we chose the one which incorporated over-dispersion (in the documentation, this is called \textit{Negative Binomial}, as the mean-variance relationship is the same as in a Negative Binomial model):
\begin{equation*}
I_{t}|\iota_{t} \sim \text{Normal}\left(\iota, \iota + \frac{\iota^{2}}{\phi}\right).
\end{equation*}
We used the default setting, which fixed the $\frac{1}{\sqrt{\phi}}=0.1$. 
Next we define 
$\lambda_{t} = \sum_{s = 0}^{L} I_{t-s} \tau_{s}^{inc}$
to be the collection of individuals actively shedding. 
In our case we used the setting \textit{from symptom onset}, but provided the distribution of the time to becoming infectious, which is when shedding begins in our simulation. 
There are two options now for defining expected shedding. 
Without individual variation:
\begin{equation*}
\omega_{t} = \sum_{s}^{S}\lambda_{t-s}\mu^{load}\tau_{s}^{shed}.
\end{equation*}
So that $\mu^{load}$ is the expected shedding load per person, and $\tau^{shed}$ is the shedding load profile. 
This is closer to what the EI-ww models do. 
The other option is closer to the truth, which is to say that 
\begin{equation*}
\omega_{t} = \sum_{s}^{S}\zeta_{t-s}\mu^{load}\tau_{s}^{shed}
\end{equation*}
where 
\begin{equation*}
\zeta_{t} \sim \text{Gamma}(\lambda_{t}/\nu^{2}, \frac{1}{\mu^{load}\nu^{2}})
\end{equation*}
which is approximated by an unspecified Normal approximation in the actual code. 
This allows for some variation in shedding for incidences. 
We chose the latter option.
Finally, we use the constant noise model so that the observed concentration $X_{t,i}$ is modeled as 
\begin{equation*}
X_{t,i} \sim \text{Log-Normal}(\log{\omega_{t}}, \sigma_{\text{noise}}).
\end{equation*}
The effective reproduction number is modeled with a random walk on a daily time scale.
At time of writing, the time scale cannot be changed. 
We chose to use the inverse soft-plus link function, so that 
\begin{equation*}
\frac{\log{1 + e^{k R_{t}}}}{k} \sim \text{Normal}\left(\frac{\log{1 + e^{k R_{t-1}}}}{k}, \sigma_{rw}\right). 
\end{equation*}
We used the default value of $k = 4$. 
\subsection{Simulation Protocol}\label{sec:sim}
\subsubsection{Simulation Engine} \label{sec:sei7r}
An agent-based stochastic SEIIIIIIIRRR model is an N-dimensional continuous time Markov chain, where N is the population. 
When represented as a vector $\mathbf{G}(t)$, each entry of $\mathbf{G}(t)$ records the state of one of the $N$ individuals, i.e. if the $ith$ entry of $\mathbf{G}(t)$, $\mathbf{G}(t)_{i}$ is $S$, then the $ith$ individual is susceptible at time $t$. 
Define $I(t) = \sum_{j=1}^{7}I_{j}(t)$ be the total number of individuals infections at time $t$.
It can be defined in terms of its transition rates from state $\mathbf{G}$ to state $\mathbf{G}'$ so that
\begin{equation*}
\lambda_{\mathbf{G}\mathbf{G}'} =
    \begin{cases}
        \beta_{t}/N \times I(t) \text{ if $\mathbf{G}_{j} = S$ and $\mathbf{G}'_{j} = E$} \\
        \gamma \text{ if $\mathbf{G}_{j} = E$ and $\mathbf{G}'_{j} = I$} \\
        \nu \text{ if $\mathbf{G}_{j} = I_k$ and $\mathbf{G}'_{j} = I_{k +1}$ for $k = 1, \dots, 6$ or if $\mathbf{G}_{j} = I_7$ and $\mathbf{G}'_{j} = R_{1}$}  \\
        \eta \text{ if $\mathbf{G}_{j} = R_{l}$ and $\mathbf{G}'_{j} = R_{l + 1}$ for $l = 1,2$} \\
        0 \text{ otherwise}
    \end{cases}
\end{equation*}
We use the well known Gillespie algorithm popularized in \citep{gillespie1977exact} to simulate realizations of this model
\par 
In order to calculate individual genome concentrations, we constructed a piecewise constant mean shedding load profile, so that, for instance, all individuals in the $I1$ compartment shed on average the same amount. 
The mean shedding in each compartment is visualized in Appendix Figure \ref{fig:avg_shedding}
\begin{figure}[H]
    \centering
    \includegraphics[width=\textwidth]{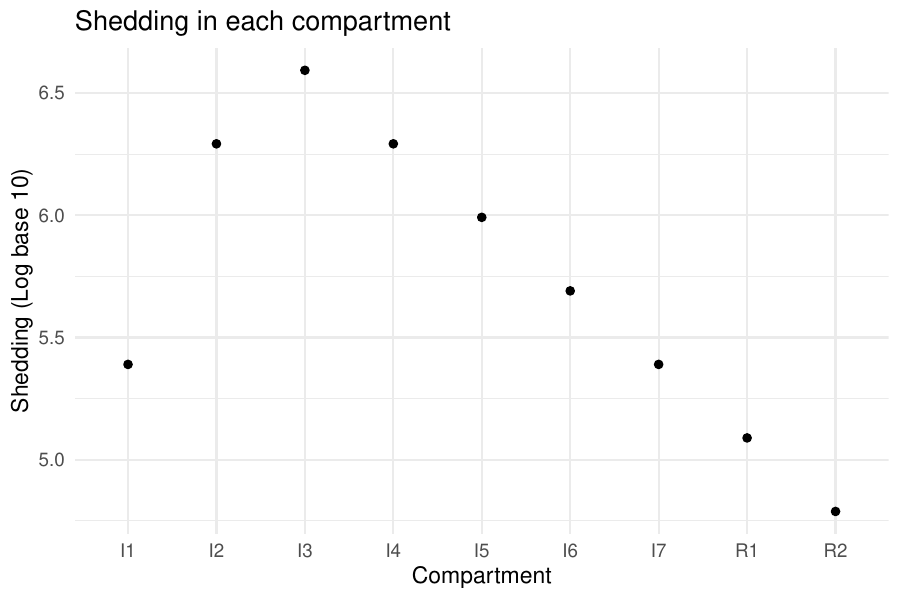}
    \caption{Mean shedding for individuals in the seven infectious states and first two recovered states.}
    \label{fig:avg_shedding}
\end{figure}
This profile was loosely based on the consensus shedding profile for SARS-CoV-2 RNA created by \citet{nourbakhsh2022wastewater}. 
Next each individual was assigned an individual random intercept on the log base 10 scale from a normal distribution with mean zero, and standard deviation 1.09 (the value 1.09 was based on the heterogeneity of shedding SARS-CoV-2 shown in the studies of \cite{hoffmann2021faecal}, which uses data collected by \citet{wolfel2020virological}, \citet{han2020viral}, and \citet{lui2020viral}). 
Population level shedding was the mean shedding of all individuals in the population at any given time. 
Observed concentrations were then modeled as log normal random variables where $X_{t} \sim \text{Log Normal}(\log{(\text{Population shedding at time t})} \times \rho, \tau)$, where $\rho$ was chosen arbitrarily to make the scale of the data match the scale of real data, and $\tau$ was taken from a previous study \citep{goldstein2024semiparametric}. 
\subsubsection{Simulation parameters}\label{sec:sim_params}
We used parameters following the logic described in \cite{goldstein2024semiparametric}. 
For convenience, we include a lightly modified version of the supplemental section from that paper which describes our approach. 
\begin{table}[H]
    \centering
    \begin{tabular}{ccc}
         Parameter & Interpretation & Value\\
           \hline \\
           $1/\gamma$ & Mean latent period duration & 4\\
           $1/\nu$ & One seventh of the mean infectious period & 1\\
           $1/\eta$ & One half of the duration when recovered but still shedding RNA & 9 \\
           $\rho$ & scales concentrations of RNA into observed concentrations & $0.0009$\\
           $\tau$ & describes the noisiness of observed gene data & 0.5 
\end{tabular}
\caption{Simulation parameters used in the simulation study. Durations are measured in days.}
\label{tbl:simsetting}
\end{table}
The estimate for $1/\gamma$ was calculated by averaging the mean latent period calculated by \citet{Xin2021} with the mean time to detecting virus which could be cultured found in \citep{killingley2022safety}. 
We took culturable virus to be a proxy for infectiousness.  
The mean time from infectiousness to symptom onset was 1.37, calculated again as an average from the previous two studies (1.4 days from \citep{Xin2021} versus 1.33 from \citep{killingley2022safety}). 
Mean infectious period was calculated using the mean period of detecting virus which could be cultured found in \citep{killingley2022safety}. 
Many studies have calculated the time from symptom onset to the end of RNA shedding in fecal matter; we averaged the estimates from \citep{okita2022duration} and \citep{zhang2021prevalence}. 
Because these two literature reviews shared studies, we dis-aggregated their estimates into individual study estimates, counting each study only once in our final average. We used mean shedding estimates from each study reported by \citet{okita2022duration}. 
\citet{zhang2021prevalence} did not include estimates of the mean for each study in their review, if estimates of the mean duration were available from the original paper, they were used, if they were not, the paper was not included in the final average. 
We also examined the literature review by \citet{walsh2020sars}, but found no new studies with more than two samples not in the previous two reviews. 
We decided to exclude studies with fewer than three samples. 
The final mean duration from symptom onset to the end of RNA shedding in fecal matter was 22.99 days. 
We calculated $1/\eta$ (the mean duration of shedding after recovery) as 
\[
2 * 1/\eta = 22.99 + 1.33 - 7 * 1/\nu = 17.86.
\]
Note we are assuming shedding begins at the start of the infectious period. 
Note that all of the parameters are based on studies of the original Wuhan lineage of SARS-CoV-2. 
To calculate a plausible $\tau$ (standard deviation from true genetic concentration), we fit a Bayesian thin plate regression generalized student t-distribution spline to wastewater data from Los Angeles, California. 
We used the mean of the posterior estimate of $\tau$ for the $\tau$ in our simulation. 

\subsubsection{Example Simulation}
 For reference, we include the shedding during the infectious stage of 9 individuals simulated using the simulation engine, as well as a visualization of an example simulation. 
\begin{figure}[H]
    \centering
    \includegraphics[width=1.0\textwidth]{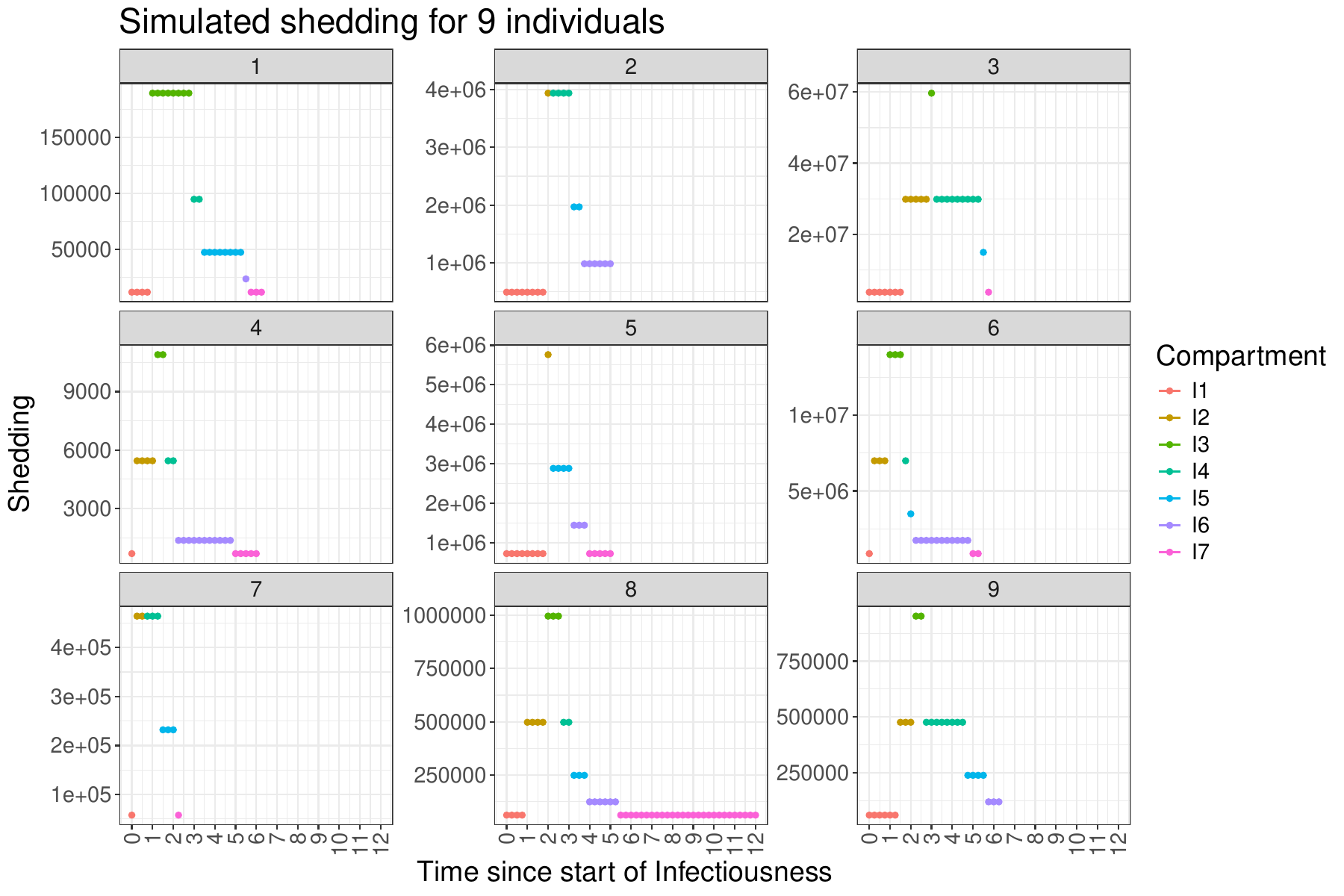}
    \caption{Simulated shedding of nine individuals. Each panel is an individual. Colors represent the the compartment the individual was in while shedding. Individuals shed the same amount in each compartment on average, but the exact amount is modified by a random intercept to incorporate heterogeneous shedding amongst individuals. The amount of time spent in each compartment is likewise random, which changes the overall shedding profile.}
    \label{fig:example_shedding}
\end{figure}

\begin{figure}[H]
    \centering
    \includegraphics[width=1.0\textwidth]{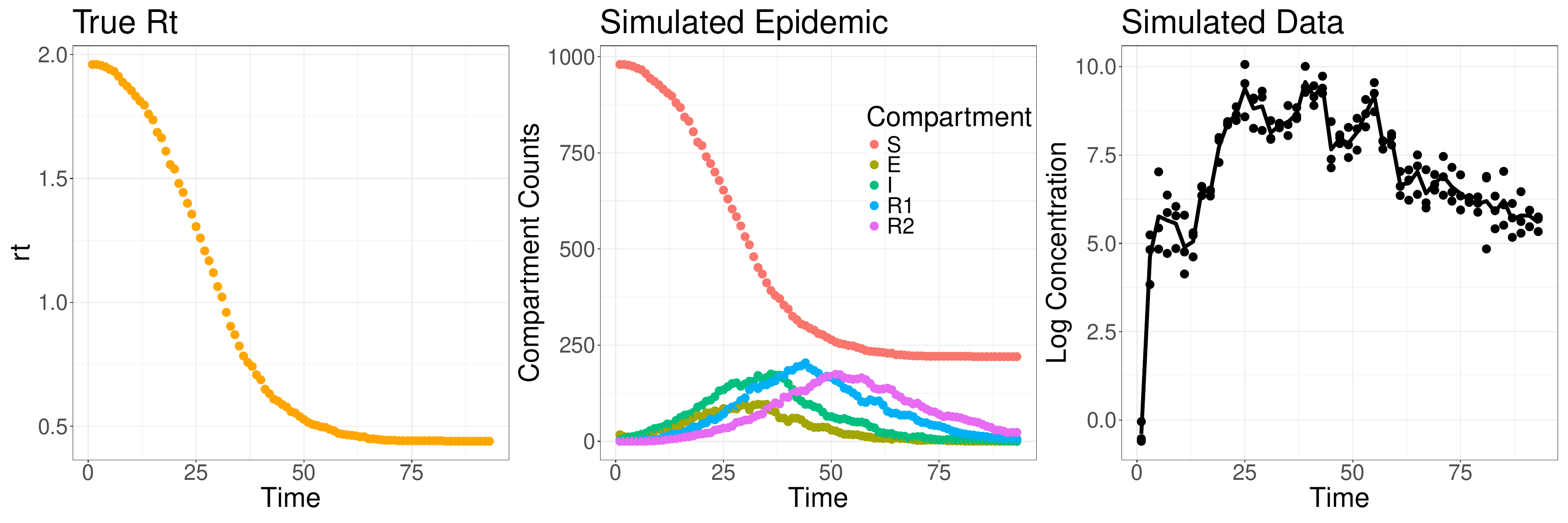}
    \caption{Simulated epidemic and corresponding wastewater data for the Fixed Rt scenario.
    The left panel displays $R_{t}$, the middle panel shows the counts of individuals in each stage of infection, and the right panel shows the wastewater data simulated from this epidemic.
    For visualization purposes, the seven I compartments are added together into one compartment "I" representing all currently infectious individuals.
    In the third panel, the dots are replicates, while the black line represents the mean of the three replicates on the log scale.}
    \label{fig:sim_data}
\end{figure}

\subsubsection{Simulation Priors}\label{sec:sim_priors}
\begin{table}[H]
\caption{Priors used by all models in the Steep simulation scenario.}
\centering
\small
\fbox{%
\begin{tabular}{*{5}{c}}
          Parameter & Model & Prior & Prior Median (95\% Interval) & Truth \\
         \hline \\
         $\gamma$ & All & Log-normal($\log(1/4)$, 0.2) & 0.25 (0.17, 0.37) & 0.25 \\
         $\nu$ &  All &  Log-normal($\log(1/7)$, 0.2) & 0.14 (0.10, 0.21) & 0.14  \\
         $\sigma_{rw}$ & All & Log-normal(log(0.2), 0.1) & 0.2 (0.16, 0.24) & NA \\
         $\tau$ & All &  Log-normal(0, 1) & 1.00 (0.14, 7.10) & 0.5\\
         $\rho$ & All & Log-normal(0, 1)  & 1.00 (0.14, 7.10) & NA \\
         $R_{0}$ &  All & Log-Normal(log(1.0), 0.1) & 1.0 (0.9, 0.1.33) & 1.1 \\
         $E(0)$ & All & Log-Normal(log(20), 0.05) & 20 (7.51, 53.29) & 20 \\
         $I(0)$ & All & Log-Normal(log(1), 0.05) & 0 (2.67) & 0 \\
\end{tabular}}
\label{tab:stoch_ww_sim_priors}
\end{table}
For the Fixed scenario, we centered the prior on the initial effective reproduction number at 1.9 rather than 1.0, but the standard deviation parameter remained the same. 
Similarly, the priors on initial compartment counts were shifted to 10 or 40 for the total500 and total2000 scenarios. 
\subsubsection{Episewer Parameters and Priors}\label{sec:episewer_priors}
The Episewer shedding load profile is a gamma distribution. 
We found shape and scale parameters for the distribution by minimizing the squared difference between the normalized points in the shedding profile in Figure A2 and the pdf of a gamma distribution.
The shape parameter was $\alpha = 6.44$ and the scale parameter was $\beta = 2.26$.
The incubation distribution was taken to be the latent period distribution, exponential with mean 4. 
The generation distribution is parametrized using the mean and standard deviation of a gamma distribution. 
We chose to empirically estimate the mean and standard deviation by simulating 10000 latent periods from an exponential distribution with mean 4, and 10000 infectious periods from a gamma distribution with shape 7 and rate 1. 
The mean used for fitting Episewer to simulations was 11.02 with standard deviation 4.86. 
\par 
For the initial prior on $R_{t}$, we minimized the squared loss of the 95\% quantiles on the prior used for the Stochastic EI model, and the initial distribution in Episewer (where the softplus of $R_{t}$ is Normal).
For the Fixed scenario the mean was $\mu = 1.9$ with standard deviation  was $\sigma = 0.19$.
For the Steep scenario the mean was $\mu = 1.02$ with standard deviation  was $\sigma = 0.1$.
We attempted to use similar techniques to try and choose a prior for the random walk standard deviation that would align the prior as closely to our own random walk prior as possible.
However, we found the default choices in Episewer led to better model performance (analysis not shown), and so we used those. 
In this case, $\mu_{\sigma_{rw}} = 0$, $\sigma_{\sigma_{rw}} = 0.1$.
\par
In the case of the real data analysis, we used a shedding load profile derived from the work of \citet{nourbakhsh2022wastewater} which we used when testing another branching process method in a previous analysis \citep{goldstein2024semiparametric}.
For the real data $\alpha = 2.18$, $\beta = 1.84$.
The generation time distribution we used comes from the work of \citet{sender2022unmitigated}, and parametrizes the standard deviation following the conventions of \citet{cori_new_2013}.
We set the prior mean for $R_{t}$ to be 0.5, with the prior standard deviation of 0.1, in order to closely match the priors of our other models.

\subsection{Calculating Initial Conditions for Real Data}
We used case data to create a rough guess for the initial conditions for our stochastic and deterministic EI-ww models. 
We assumed the total number of individuals in the $E$ and $I$ compartments was equal to the last 11 days before the start of the observation period (the sum of the mean latent period and mean infectious period) of reported cases multiplied by 2 (i.e. an under-reporting rate of 0.5). We then split this total so that two thirds went to the $I$ compartment and one third went to the $E$ compartment. 
These case data were available at the community level, we then split the calculate counts proportionally by estimates of the proportion of the community living in each sub-community. 
For the initial effective reproduction number, we chose a prior centered around 0.5, this was based on previous estimates of the effective reproduction number during this time using case data \citep{goldstein2024incorporating}.
The final priors are displayed in the table below. 
\begin{table}[H]
\caption{EI-ww priors for UC Irvine.}
\centering
\fbox{%
\begin{tabular}{*{5}{c}}
          Parameter & Place &  Prior & Prior Median (95\% Interval) \\
         \hline \\
         $E(0)$ & G1 &  Log-Normal(log(3), 0.05) & 3 (2.72, 3.31) \\
         $I(0)$ & G1 &  Log-Normal(log(5), 0.05) & 5 (4.53, 5.51)  \\
         $E(0)$ & G2 &  Log-Normal(log(3), 0.05) & 3 (2.72, 3.31) \\
         $I(0)$ & G2 &  Log-Normal(log(6), 0.05) & 6 (5.44, 6.62)  \\
         $E(0)$ & E1 &  Log-Normal(log(2), 0.05) & 2 (1.81, 2.21) \\
         $I(0)$ & E1 &  Log-Normal(log(4), 0.05) & 4 (3.63, 4.41)  \\
         $R_{0}$ & All & Log-Normal(log(0.5), 0.1) & 2 (1.64, 2.43)
\end{tabular}}
\label{tab:uci_init_conds}
\end{table}
\subsection{Priors and Posteriors of Fixed Parameters}
We provide a visualization of the priors and posteriors of the fixed parameters in the stochastic EI-ww model for the model fit to the Fixed scenario visualized in Figure 1 of the main text. 
We also show the prior and posterior of $\tau$ in particular, as the scale is hard to see in the original figure. 

\begin{figure}[H]
    \centering
    \includegraphics[width = \textwidth]{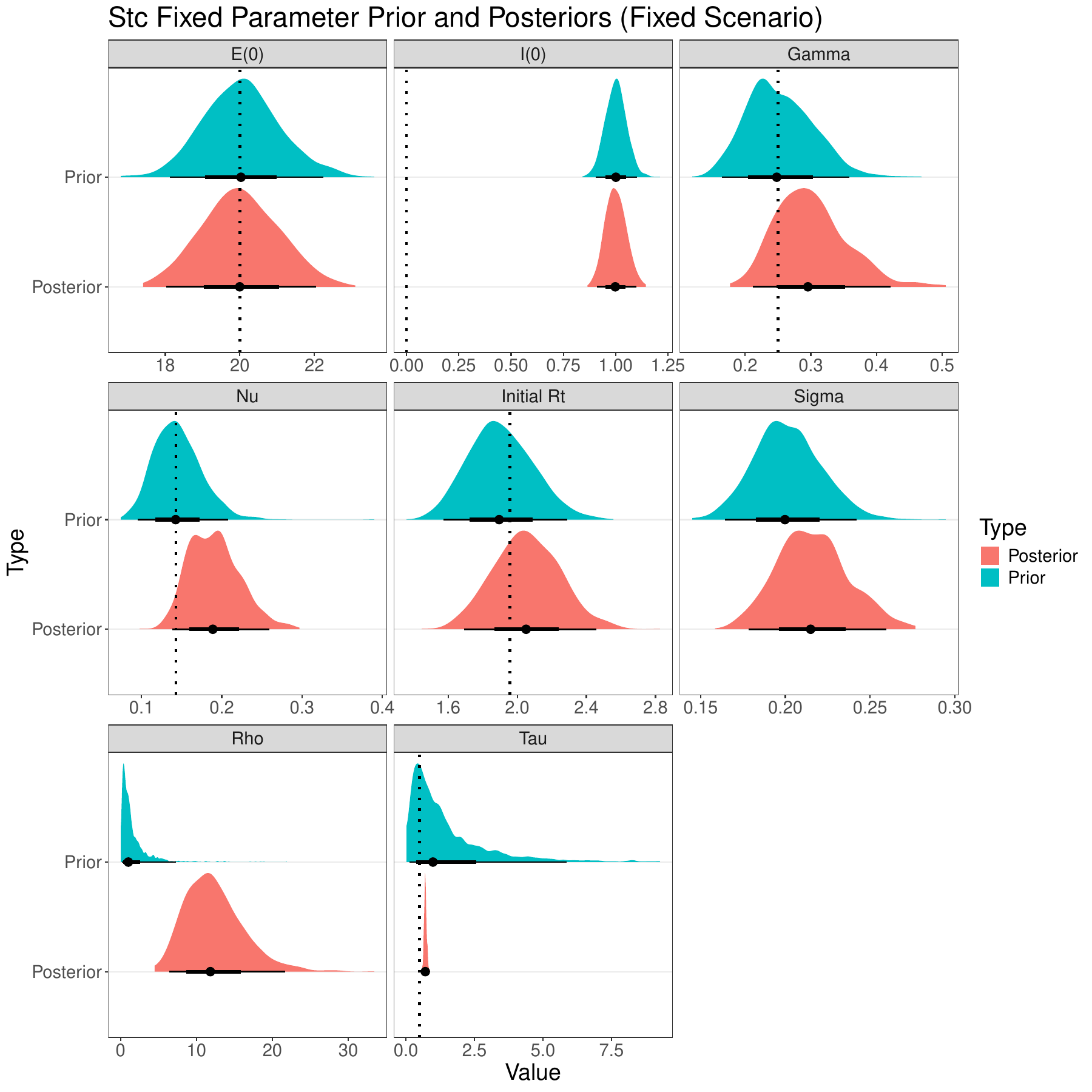}
    \caption{Stochastic EI-ww prior and posterior densities for fixed model parameters.
        Posterior summaries are from the model fit to the Fixed Scenario data shown in Figure 1 of the main text. 
    Blue densities are the prior, red densities are the posterior, dotted lines indicate true values (when relevant).}
    \label{fig:ch5_example_fixed_pp}
\end{figure}

\begin{figure}[H]
    \centering
    \includegraphics[width = \textwidth]{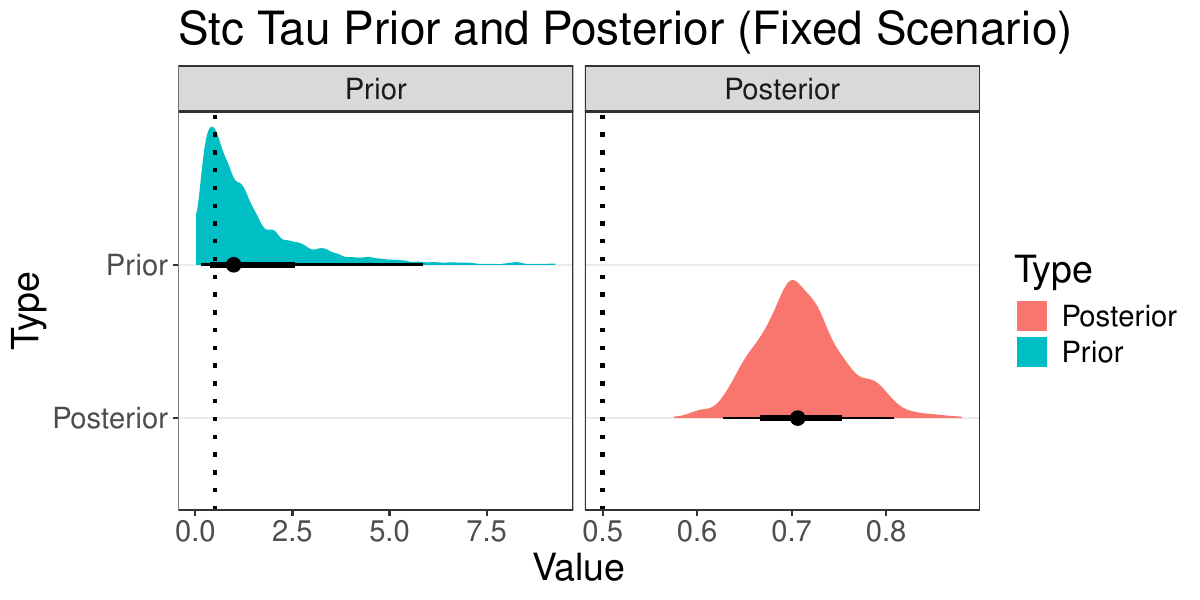}
    \caption{Stochastic EI-ww prior and posterior densities for fixed model parameters for the parameter $\tau$. 
    The figure is a transformation of the bottom right panel of Figure \ref{fig:ch5_example_fixed_pp}.}
    \label{fig:ch5_example_fixed_taupp}
\end{figure}
\section{Posterior Predictive Real Data Analysis}
\begin{figure}[H]
    \centering
    \includegraphics[width =\textwidth]{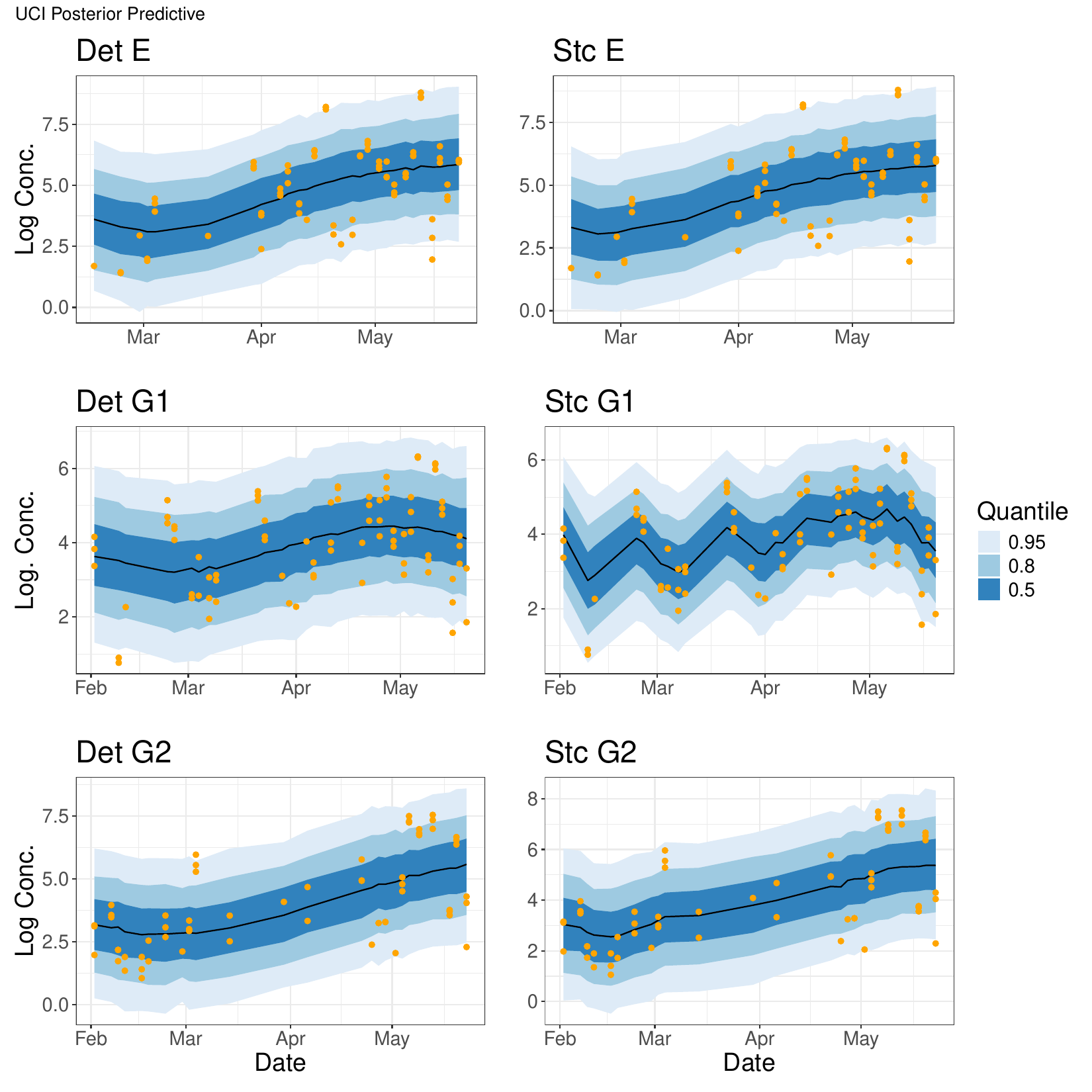}
    \caption{Posterior predictive summaries from UCI wastewater data. 
    Real data are orange dots, the posterior predictive median is the black line, blue regions are distribution quantiles of varying width.}
    \label{fig:uci_postpred}
\end{figure}

\section{Real Data Sensitivity Analysis}
\begin{figure}[H]
    \centering
    \includegraphics[width = \textwidth]{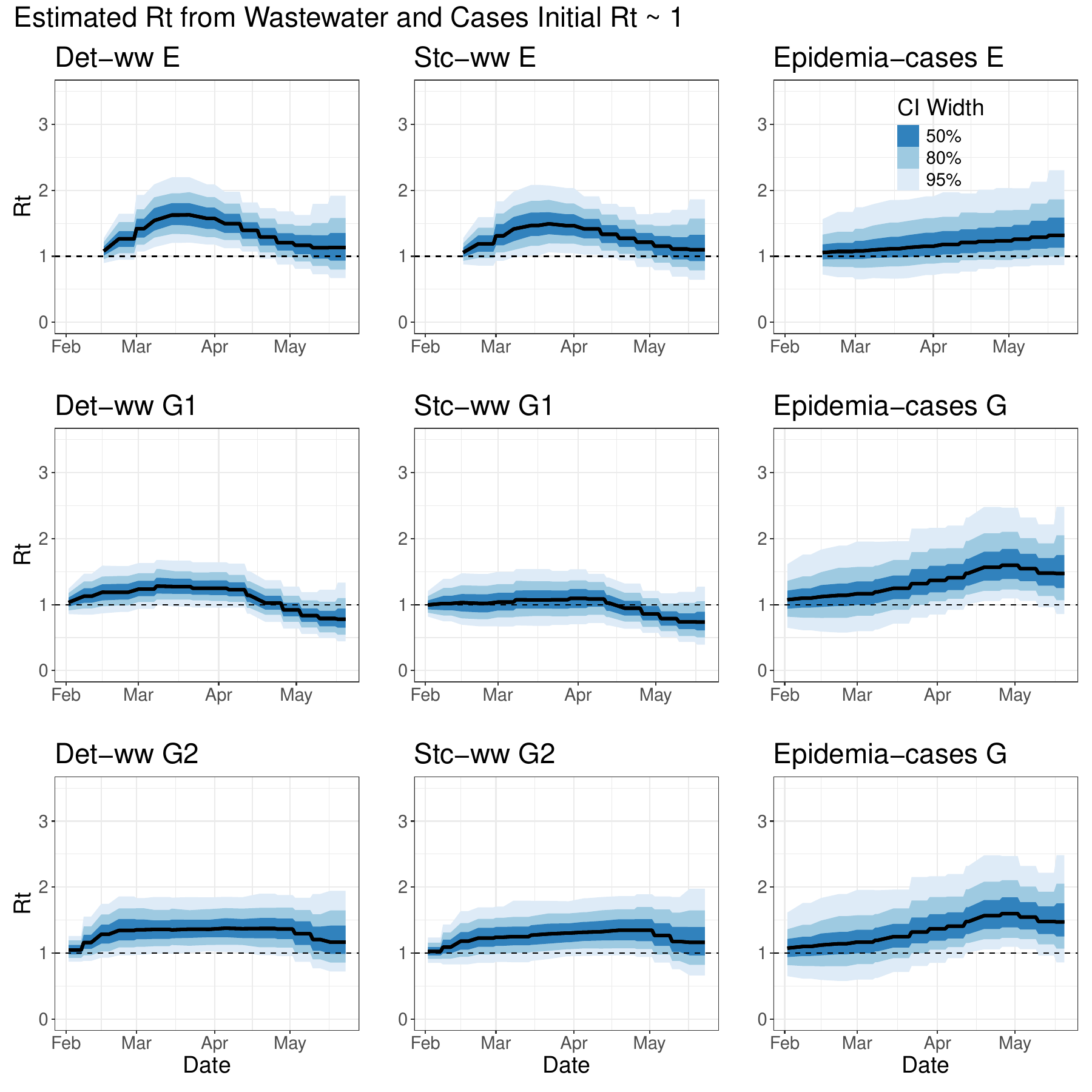}
    \caption{Posterior summaries of the effective reproduction number in college campus communities estimated from wastewater data and case data using a prior on the initial $R_{t}$ centered at 1 rather than 0.5. Black lines are medians, blue shaded regions are credible intervals.}
    \label{fig:uci_rt_rt=1}
\end{figure}

In the main analysis we used an initial $R_{t}$ derived from previous case-based estimates of $R_{t}$ for the entirety of Orange County, CA. 
It is plausible this county-wide $R_{t}$ did not accurately reflect the situation at UC Irvine at the start of the modeling period. 
To understand how our choice of prior affected our analysis, we re-analyzed the data using an alternative prior centered around $1$ instead of $0.5$ ($R_{0} \sim \text{Log-Normal}(0, 0.1)$). 
Figure \ref{fig:uci_rt_rt=1} displays the results of this re-analysis. 
Many of the patterns we observed in the main analysis are still present. 
The wastewater model results are quite different from case model results.
The deterministic EI-ww model is a little less certain than in the main analysis, but still more certain overall than the stochastic EI-ww model.
In the case of the E community, the stochastic model now has 95\% credible interval above 1 for a brief time period, while the case model never has 95\% credible intervals above 1. 
\end{document}